\documentclass[reprint,amsmath,amssymb,aps,pra,reprint,
 amsmath,amssymb,
 pra,superscriptaddress]{revtex4-2}

\usepackage{mypackages}
\usepackage{amscd}
\usepackage{textcomp}
\usepackage[thinlines]{easybmat}
\usepackage{multirow,bigdelim}

\usepackage{cleveref}
\usepackage{xcolor} 
\usepackage{soul}
\begin{document}

\title{Fault-Tolerant Logical Clifford Gates from Code Automorphisms}

\author{Hasan Sayginel}
\thanks{\ These authors contributed equally to this work. Corresponding author: \url{hasan.sayginel.17@ucl.ac.uk}}
\affiliation{Department of Physics \& Astronomy, University College London, London, WC1E 6BT, United Kingdom}
\affiliation{National Physical Laboratory, Teddington, TW11 0LW, United Kingdom}
\author{Stergios Koutsioumpas}
\thanks{\ These authors contributed equally to this work. Corresponding author: \url{hasan.sayginel.17@ucl.ac.uk}}
\affiliation{Department of Physics \& Astronomy, University College London, London, WC1E 6BT, United Kingdom}
\author{Mark Webster}
\thanks{\ These authors contributed equally to this work. Corresponding author: \url{hasan.sayginel.17@ucl.ac.uk}}
\affiliation{Department of Physics \& Astronomy, University College London, London, WC1E 6BT, United Kingdom}
\author{Abhishek Rajput}
\affiliation{Department of Computer Science, University College London, London, WC1E 6BT, United Kingdom}
\author{Dan E Browne}
\affiliation{Department of Physics \& Astronomy, University College London, London, WC1E 6BT, United Kingdom}



\date{\today}

\begin{abstract}
We study the implementation of fault-tolerant logical Clifford gates on stabilizer quantum error correcting codes based on their symmetries.  Our approach is to map the stabilizer code to a binary linear code, compute its automorphism group, and impose constraints based on the Clifford operators permitted. 
We provide a rigorous formulation of the method for finding automorphisms of stabilizer codes and generalize the $ZX$-dualities previously introduced for CSS codes to non-CSS codes. 
We provide a Python package implementing our algorithms which uses the computational algebra system MAGMA for certain subroutines. 
Our algorithms map automorphism group generators to physical circuits, calculate Pauli corrections based on the destabilizers of the code, and determine their logical action. 
We discuss the fault tolerance of the circuits and include examples of gates through automorphisms for the $[[4,2,2]]$ and perfect $[[5,1,3]]$ codes, bivariate bicycle codes, and the best known distance codes. \end{abstract}

\maketitle


\section{\label{sec:introduction}Introduction}
Major obstacles to the construction of reliable large-scale quantum computers are the noise and errors in the qubits, gates and measurements which make up their implementation. There has thus been significant research interest in the development of  fault-tolerant quantum computation techniques which would allow reliable computation in the presence of hardware errors. \par

Quantum error correction is one of the most promising approaches towards fault-tolerant quantum computation, where errors are detected and corrected by encoding the information across multiple physical qubits. 

Protecting the encoded information is not enough however, and it is crucial that our codes allow for efficient implementation of logical operators on the encoded information for the algorithm we aim to perform. In particular, low depth and low resource overhead implementations of logical gates are highly desirable.\par

Transversal logical operators are an example of fault-tolerant gates \cite{gottesman97}. In transversal logical gates, no physical operation acts upon more than one qubit in a code block. This therefore avoids multi-qubit errors within the code block which could effectively reduce its distance. 
Most stabilizer codes only admit a very limited set of transversal logical operators and it has been shown that it is impossible for a code to have a universal set of transversal logical gates \cite{eastin_restrictions_2009}.



The notion of transversality can be slightly relaxed to  \textbf{\textit{SWAP}-transversal} logical gates which include both qubit permutations as well as single-qubit Clifford gates. 
For qubit architectures where physical \textit{SWAP} operations can be implemented with low overhead, this allows for a wider range of fault-tolerant gates.

\textit{SWAP}-transversal gates of a stabilizer code can be identified by looking at the automorphisms of a related binary linear code.
For binary linear codes, the term \textbf{automorphism}  denotes a permutation of the bits of a code that map a code to itself. 
The automorphisms of a code form a group under composition. Calderbank et al.\ \cite{calderbank1997}, extended the notion of automorphism to quantum codes by showing that quantum codes can be represented as classical codes over $GF(4)$. 
 In this representation, automorphisms of the classical code correspond to physical qubit permutations and single-qubit Clifford gates which preserve the codespace and implement fault-tolerant logical operations.

 

 
In \cite{moussa_transversal_2016,breuckmann_fold-transversal_2022,Quintavalle2023partitioningqubits}, automorphisms of the binary linear codes used in the construction of a CSS code were used to identify \textbf{fold-transversal} gates which employ  qubit permutations and Hadamard ($H$) gates on all physical qubits. 
 
In this paper we bridge the gap between the fold-transversal and $GF(4)$ methods by introducing a family of algorithms for finding fault-tolerant Clifford gate implementations for any stabilizer code based on symmetries of the code. 
By choosing different binary representations of stabilizer codes, we can control which single-qubit Clifford operations appear in the circuits implementing a logical gate.
This generalizes the $ZX$-dualities of \cite{breuckmann_fold-transversal_2022} to non-CSS codes and a wider range of logical gates.
We develop a rigorous approach for finding Clifford gates through code automorphisms that includes applying Pauli corrections to ensure the stabilizer group is preserved and identifying logical actions and physical circuits.  \par

We provide a package with an implementation of our algorithms in Python using MAGMA \cite{bosma_magma_1997} and the open-source Bliss \cite{Bliss} package for the calculation of the automorphism group of a related classical linear code. The package is available at \url{https://github.com/hsayginel/autqec}.\par

The structure of the paper is as follows. In  \Cref{sec:backgroud} we introduce the key theoretical concepts outlined above. We then discuss stabilizer codes, logical operators and fault tolerance. 
In \Cref{sec:FaultTolerance} we discuss the fault tolerance of the circuits we obtain by separating them in three different categories. We highlight the close link between the hardware implementation of a \textit{SWAP} gate and the fault-tolerant properties of our circuits. \par

In  \Cref{sec:auts_of_stab_codes}, we introduce our algorithms. Our methods work by mapping a stabilizer code to a related binary linear code with constraints on the physical Clifford gates included in the circuits we want to find. 
We first show how to obtain the logical operators consisting of physical \textit{SWAP} and $H$ gates for general stabilizer codes. 
These arise as a subgroup of the automorphisms of the classical linear code generated by the quantum code's symplectic representation. 
In the case of CSS codes, we recover the $ZX$-dualities from \cite{breuckmann_fold-transversal_2022} but also find a wider range of operators. 
Similarly, we can fix any of the other two Pauli axes and obtain circuits consisting of physical \textit{SWAP} and transversal $S$ or $\sqrt{X}$ gates. \par

We then extend this method by using a $3-$block representation which recovers the form of \cite{calderbank1997} and allows us to find logical operators consisting of physical \textit{SWAP} and arbitrary single-qubit Clifford gates. We expand on the method by finding Pauli phase corrections and determining the logical action of the circuit. \par 

We then explain a number of technical points for the methods. 
In \Cref{sec:pauli_correction}, we show how to apply Pauli corrections to automorphism gates to ensure that the signs of the stabilizer generators are preserved. 
In \Cref{sec:logical_action}, we show how to find the logical action of the gates.
In \Cref{sec:sym2circuits}, we show how to translate the results of the algorithms to a standard circuit form.

In \Cref{sec:embedded}, we show how to use the embedded code technique of \cite{webster_transversal_2023} to map pairs of qubits onto additional columns of the related binary linear code.
This allows us to represent physical \textit{CNOT} and \textit{CZ} gates as automorphisms. 
We show that different embedding operators allow us to restrict the physical circuits we find based on additional constraints. This can be used to find a wider range of Clifford logical actions and to restrict \textit{CNOT} and \textit{CZ} gates in accordance with the connectivity of the available hardware.\par

In \Cref{sec:examples_and_results}, we apply our techniques to various stabilizer codes.
We first find fault-tolerant gates for the well-known $[[4,2,2]]$ and $[[5,1,3]]$  codes. 
We then analyze the automorphism groups of the best-known-distance single logical qubit stabilizer codes for up to $30$ physical qubits from  \cite{Grassl:codetables}.
Finally, we calculate and analyze the automorphism groups of the bivariate bicycle codes of \cite{bravyi_high-threshold_2024} and find additional automorphism gates compared to that work. 

In \Cref{sec:aut_complexity}, we discuss the computational complexity of our algorithms.

\begin{figure}[ht]
\includegraphics[width=0.45\textwidth]{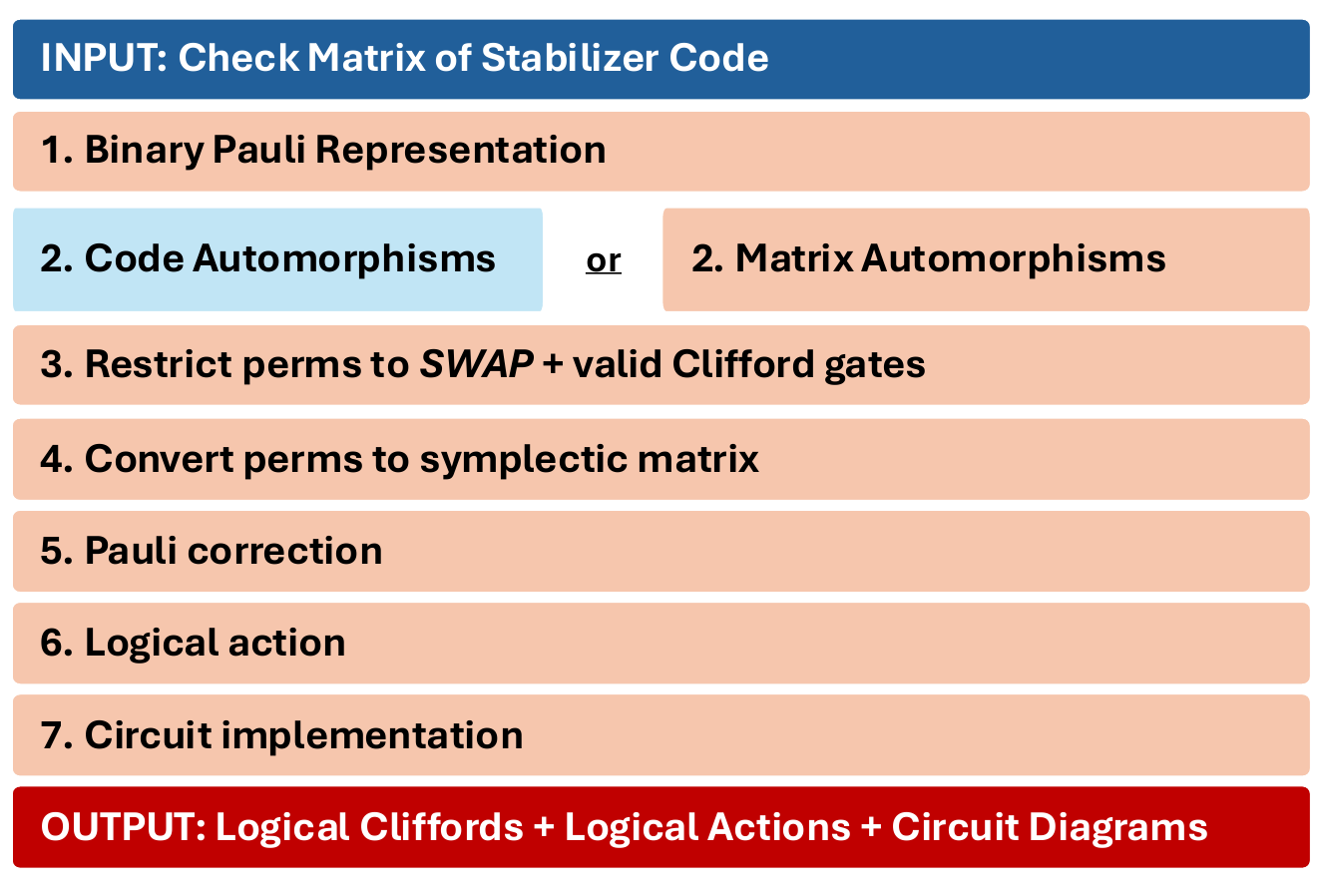}
\caption{Outline of our algorithms for finding logical Clifford operators via automorphisms of a related binary linear code.
The step outlined in light blue (code automorphisms) is performed using a computational algebra package such as MAGMA. All other steps are performed in Python.}
\label{fig:algo_outline}
\end{figure}

\Cref{fig:algo_outline} outlines our set of algorithms: Step 1 is explained in \Cref{sec:backgroud} for single-qubit gates and in \Cref{sec:embedded} for two-qubit gates. Steps 2-4 are explained in \Cref{sec:auts_of_stab_codes}. Step 5 is explained in \Cref{sec:pauli_correction}. Step 6 is explained in \Cref{sec:logical_action} and Step 7 is explained in \Cref{sec:sym2circuits}.

\section{Background}\label{sec:backgroud}

\subsection{Stabilizer Codes}\label{sec:stabilizer_codes}
In this section, we introduce the key concepts of stabilizer codes and illustrate these using the well-known 5-qubit code as an example. 
Stabilizer codes are defined by choosing a set $\mathbf{G}$ of \textbf{stabilizer generators.} 
These are a set of independent commuting Pauli operators on $n$ qubits which generate the \textbf{stabilizer group} denoted $\braket{\mathbf{G}}$. 
The \textbf{codespace} is the subspace of states on $n$ qubits $\mathcal{H}_2^n$ fixed by all elements of $\mathbf{G}$ - that is the set $\{\ket{\psi} \in \mathcal{H}_2^n : A\ket{\psi} = \ket{\psi}, \forall A \in \mathbf{G}\}$. 
\subsubsection{Symplectic Representation of Pauli Operators}
In this work, we use various mappings of the stabilizer generators of a quantum code to a binary matrix which can then be interpreted as the generator matrix of a binary linear code.
Pauli operators on $n$ qubits have a \textbf{symplectic representation} as binary vectors of length $2n$. 
Let $\mathbf{x, z}$ be binary vectors of length $n$. We define an \textbf{unsigned Pauli operator} as:
\begin{equation}
X(\mathbf{x}) Z(\mathbf{z}) : = \prod_{0 \le i <n}X_i^{\mathbf{x}[i]} \prod_{0 \le i <n}Z_i^{\mathbf{z}[i]}.
\end{equation}
We refer to the vector $\mathbf{x}$ as the \textbf{X-component} and $\mathbf{z}$ as the \textbf{Z-component} of the operator.
The operators of form $X(\mathbf{x}) Z(\mathbf{z})$ generate the Pauli group on $n$ qubits, up to global phases. 
The single-qubit Pauli operators have the following binary string representations:
\begin{align*}
    X \equiv (1|0);\; Z \equiv (0|1);\; Y = iXZ \equiv (1|1).
\end{align*}
Note that in the case of the $Y$ operator, the representation does not take into account the global phase $i$.
Up to global phases, multiplication of Pauli operators corresponds to addition of the vector representations modulo $2$. 

We can check whether two Pauli operators commute via the \textbf{symplectic inner product} of their vector representations\textbf{.} The Pauli operators $X(\mathbf{x}_1) Z(\mathbf{z}_1)$ and $X(\mathbf{x}_2) Z(\mathbf{z}_2)$ commute if and only if:
\begin{align}
(\mathbf{x}_1,\mathbf{z}_1) \Omega_n (\mathbf{x}_2,\mathbf{z}_2)^T = \mathbf{z}_1 \cdot \mathbf{x}_2 +\mathbf{x}_1 \cdot \mathbf{z}_2 =   0 \mod 2, \label{eq:pauli_commutation}
\end{align}
where  $\Omega_n := \begin{bmatrix}0_{n\times n}&I_n\\I_n&0_{n\times n}\end{bmatrix}$ is the \textbf{binary symplectic form}. 
We will use $\Omega$ going forward rather than $\Omega_n$ because $n$ is determined by the number of physical qubits used by the stabilizer code.

\subsubsection{Logical Pauli Operators}
Pauli operators which commute with all elements of $\mathbf{G}$ but are not in the stabilizer group $\langle \mathbf{G}\rangle$  are called \textbf{non-trivial logical Pauli operators.} 
A generating set of the non-trivial logical Pauli operators can be determined using the method of Section 10.5.7 of \cite{nielsen_quantum_2010}.
Using Gaussian elimination and permuting the qubits, we can write the check matrix $G$ of a stabilizer code in the following \textbf{standard form}:
\begin{align}
    \begin{array}{ccc}
        \begin{array}{c} G \end{array} &= 
        \left[\begin{array}{ccc|ccc}
            I\;\;\;\;\; & A_1\;\;\;\;\; & A_2\;\; & \;\;B\;\;\;\;\; & 0\;\;\;\;\; & C_1 \\
            0\;\;\;\;\; & 0\;\;\;\;\; & 0\;\; & \;\;D\;\;\;\;\; & I\;\;\;\;\; & C_2
        \end{array}\right] &
        \begin{array}{cc}
            \rdelim\}{0.8}{1mm}[r] \\
            \rdelim\}{0.8}{1mm}[s]
        \end{array}
        \\
        & \begin{array}{cccccc}
            \;\;\;\;\underbrace{}_{r} & \underbrace{}_{s} & \underbrace{}_{k} & \;\;\;\;\underbrace{}_{r} & \underbrace{}_{s} & \underbrace{}_{k}
          \end{array} 
    \end{array}
    \label{eq:standard_form}
\end{align}
The rank of the check matrix $G$ determines the number of \textbf{logical qubits} $k := n - r - s$. 
We can construct a set of $k$ independent \textbf{logical Z} operators and $k$ independent \textbf{logical X} operators as follows:
\begin{align}
L := \begin{bmatrix} L_X \\ L_Z \end{bmatrix} = 
\left[
\begin{array}{ccc|ccc}
    0 & C_2^T & I & C_1^T & 0 & 0 \\
    0 & 0 & 0 & A_2^T & 0 & I
\end{array}
\right]
\label{eq:logical_paulis}
\end{align}
Each row of $L_Z$ anticommutes with the corresponding row of $L_X$, and commutes with all other rows.
\begin{example}\label{eg:513_standard_form}
    The stabilizer group of the perfect $[[5,1,3]]$ Quantum Error Correcting code is generated by: 
    \begin{equation}
        \text{\textbf{G}}=\langle XZZXI, IXZZX, XIXZZ, ZXIXZ \rangle.
    \end{equation}
    The check matrix in symplectic form is:
    \[
G:=\left[
\begin{array}{ccccc|ccccc}
1&0&0&1&0&0&1&1&0&0\\
0&1&0&0&1&0&0&1&1&0\\
1&0&1&0&0&0&0&0&1&1\\
0&1&0&1&0&1&0&0&0&1\\
\end{array}\label{513symp}
\right].
\]
Calculating the standard form as in \Cref{eq:standard_form} we get:
\begin{equation}\label{eq:n5code_sform}
    G =  \begin{bmatrix}g_1\\g_2\\g_3\\g_4\end{bmatrix} = \left[ \begin{array}{ccccc|ccccc}
        1&0&0&0&1&1&1&0&1&1\\
        0&1&0&0&1&0&0&1&1&0\\
        0&0&1&0&1&1&1&0&0&0\\
        0&0&0&1&1&1&0&1&1&1
        \end{array}\right]
\end{equation}
Thus its Pauli logical operator basis is:
\begin{equation}\label{eq:n5code_sform_logicals}
    \begin{bmatrix}L_X\\L_Z\end{bmatrix} = \left[ \begin{array}{ccccc|ccccc}
        0&0&0&0&1&1&0&0&1&0 \\
        0&0&0&0&0&1&1&1&1&1 
      \end{array}\right]
\end{equation}
\end{example}

\subsubsection{Representation of Clifford operators as binary symplectic matrices}
\textbf{Clifford operators} are unitaries which map Pauli operators to Pauli operators under conjugation and play a crucial role in quantum error correction. 
Our algorithms search for logical operators which are composed of physical Clifford gates and whose logical action can also be described as a Clifford unitary.
The Clifford operators form a group denoted by $C_2$ which has the following generators:
\begin{align*}
     C_2= \langle H, S, \textit{CZ} \rangle
\end{align*}
where
\begin{align*}
    H:=&\frac{1}{\sqrt{2}}\begin{bmatrix} 1 & 1 \\ 1 & -1 \end{bmatrix} \hspace{1em}
    S:=\begin{bmatrix} 1 & 0 \\ 0 & i \end{bmatrix}\\
    &\textit{CZ}:=\begin{bmatrix}
        1 & 0 &0 &0\\
        0 & 1 & 0 & 0\\
        0&0&1&0\\
        0&0&0&-1\\    \end{bmatrix} 
\end{align*}
Other important Clifford gates which we will use below are:
\begin{align*}
    \text{\textit{CNOT}}:=& \begin{bmatrix}
        1 & 0 &0 &0\\
        0 & 1 & 0 & 0\\
        0&0&0&1\\
        0&0&1&0\\\end{bmatrix}\\
    \text{\textit{SWAP}}:=& \begin{bmatrix}
        1&0&0&0\\
        0&0&1&0\\
        0&1&0&0\\
        0&0&0&1\\\end{bmatrix}\\
    \sqrt{X}:=&\frac{1}{2}\begin{bmatrix} 1+i & 1-i \\ 1-i & 1+i \end{bmatrix} \\
    C(X,X):=&\frac{1}{2}\begin{bmatrix}
        1&1&1&-1\\
        1&1&-1&1\\
        1&-1&1&1\\
        -1&1&1&1\\\end{bmatrix}\\
\end{align*}

Clifford unitaries can be represented as binary symplectic matrices up to a Pauli factor.
\textbf{Binary symplectic matrices} are binary matrices $U$ which preserve the symplectic form $U\Omega U^T = \Omega$. 
A binary symplectic matrix acting on the vector representation of a Pauli operator by right matrix multiplication has the same action as the  related Clifford operator acting via conjugation. 
We illustrate this in the following example which gives the symplectic matrix form of the $S$, \textit{CZ} and \textit{CNOT} Clifford operators:

\begin{example}\label{eg:symplectic}
    The symplectic matrix $\left[\begin{array}{c|c}1&1\\\hline0&1\end{array}\right]$ acts on single-qubit Pauli operators as the Clifford operator $S:=\begin{bmatrix}1&0\\0&i\end{bmatrix}$.
    It is sufficient to determine the action of $S$ on the basis $\{X,Z\}$ of Pauli operators on one qubit up to global phases:
    \begin{align*}
        S&: X \equiv (1|0) \mapsto Y \equiv (1|1)\\
S&: Z \equiv (0|1) \mapsto Z \equiv (0|1).
    \end{align*}
    This is equivalent to right multiplication by the matrix  $\left[\begin{array}{c|c}1&1\\\hline0&1\end{array}\right]$ as claimed.\\
Similarly, the symplectic matrix $\left[\begin{array}{c|c}0&1\\\hline1&0\end{array}\right]$ swaps the X and Z components and so acts as a $H$ operator.

The  matrix $\left[\begin{array}{cc|cc}1&0&0&1\\0&1&1&0 \\\hline0&0&1&0\\0&0&0&1\end{array}\right]$ acts on 2-qubit Paulis as an $\textit{CZ}_{01}$ operator because:\begin{align*}
\textit{CZ}_{01}&: X_0 \equiv (10|00) \mapsto X_0Z_1 \equiv (10|01)\\
\textit{CZ}_{01}&: X_1 \equiv (01|00) \mapsto X_1Z_0 \equiv (01|10)\\
\textit{CZ}_{01}&: Z_0 \equiv (00|10) \mapsto Z_0 \equiv (00|10)\\
\textit{CZ}_{01}&: Z_1 \equiv (00|01) \mapsto Z_1 \equiv (00|01).
\end{align*}
generalizing this example, symplectic matrices of form $\left[\begin{array}{c|c}I&Q\\\hline 0&I\end{array}\right]$ where $Q$ is a  symmetric binary matrix correspond to a product of $S_i$ operators (non-zero entries $Q_{ii}$ along the diagonal) and $\textit{CZ}_{ij}$ operators (non-zero entries $Q_{ij} = Q_{ji}$).\\
The binary symplectic matrix $\left[\begin{array}{cc|cc}1&1&0&0\\0&1&0&0 \\\hline0&0&1&0\\0&0&1&1 \end{array}\right]$ acts on 2-qubit Paulis as an $\textit{CNOT}_{01}$ operator because:\begin{align*}
\textit{CNOT}_{01}&: X_0 \equiv (10|00) \mapsto X_0X_1 \equiv (11|00)\\
\textit{CNOT}_{01}&: X_1 \equiv (01|00) \mapsto X_1 \equiv (01|10)\\
\textit{CNOT}_{01}&: Z_0 \equiv (00|10) \mapsto Z_0 \equiv (00|10)\\
\textit{CNOT}_{01}&: Z_1 \equiv (00|01) \mapsto Z_0Z_1 \equiv (00|11)
\end{align*}
\end{example}
generalizing this example, symplectic matrices of form $\left[\begin{array}{c|c}C&0\\\hline 0&C^{-T}\end{array}\right]$, where $C$ is an invertible binary matrix and $C^{-T}$ is the inverse of the transpose of $C$, correspond to a product of $\textit{CNOT}$ operators.\null\hfill$\Box$

\subsubsection{Encoding Operator of a Stabilizer Code}\label{sec:tableau}
Quantum information is encoded into a stabilizer code by using a Clifford operation called an \textbf{encoding operator}.
We now show how to construct an encoding operator for a stabilizer code given in terms of a check matrix $G$ in standard form and the logical Pauli generators $L_X, L_Z$.
We first construct a set of \textbf{destabilizer generators} $R$ which have the property that each row of $R$ anticommutes with the corresponding row of $G$, but commutes with all other rows of $G$, $L_Z$ and $L_X$:
\begin{align}
    R&:=\begin{bmatrix}R_X\\R_Z\end{bmatrix} = \left[\begin{array}{c c c | c c c}
      0&0&0&I&0&0\\
      0&I&0&0&0&0
    \end{array}\right]
\end{align}
We then write the stabilizers, destabilizers and logical Paulis in \textbf{tableau format} \cite{aaronson_improved_2004} as follows:
\begin{align}
\label{eq:logicalops}
\tau := 
\left[\begin{array}{c}
G_X\\G_Z\\ L_X\\\hline R_X\\R_Z\\L_Z
\end{array}\right] = 
\left[\begin{array}{c c c | c c c}
I & A_1&A_2 & B & 0 & C_1\\
0 &0&0&D&I&C_2\\
0&C_2^T&I&C_1^T&0&0\\
\hline
0&0&0&I&0&0\\
0&I&0&0&0&0\\
0&0&0&A_2^T&0&I
\end{array}\right]
\end{align}
By construction, $\tau$ is a \textbf{binary symplectic matrix} because the commutation relations of $G,L_X,R$ and $L_Z$ are equivalent to the identity $\tau\Omega \tau^T = \Omega$.

The Clifford operator $\mathcal{C}:\mathcal{H}_2^n \rightarrow \mathcal{H}_2^n$ corresponding to the binary symplectic matrix $\tau$ maps the $k$-qubit \textbf{logical state} $\ket{\psi}$ to the $n$-qubit \textbf{encoded state} $\overline{\ket{\psi}}$ as follows:
\begin{align}
\overline{\ket{\psi}} := \mathcal{C}\Big(\ket{+}^{\otimes (n-k)}\ket{\psi}\Big).
\end{align}

\subsection{\label{sec:FaultTolerance}Fault-Tolerant Logical Operators of stabilizer Codes}
The objective of this paper is to identify fault-tolerant logical Clifford operators of a given stabilizer code. 
In this section, we explain in more detail the important concepts of fault tolerance and logical operators of stabilizer codes.

\textbf{Fault Tolerance} is the property of a system that ensures that errors do not propagate catastrophically and can be systematically suppressed. 
For instance, when applying a two-qubit \textit{CZ} gate, a Pauli $X$ error on the control qubit can spread to the target qubit. 
On the other hand, applying a single-qubit $S$ gate will not inherently spread $Z$ errors.
In certain qubit architectures, \textit{SWAP} gates can be implemented with no direct interaction between qubits, meaning that the spread of Pauli errors is also constrained.
In order to quantify fault tolerance in quantum circuits, we often use the distance of the circuit, which is defined as the minimum number of faulty physical error locations that can introduce a logical error to the computation.

A \textbf{logical operator} acting on an $[[n,k]]$ stabilizer code is a unitary which maps the code subspace to itself. More specifically, a \textbf{Clifford logical operator} $\overline{U}$ is a logical operator which   applies a Clifford operator $U$ to the code subspace, transforming the logical basis $L_X,L_Z$. 

Finding a Clifford operator which applies a desired logical action is relatively straightforward. 
For instance the Clifford operator $\overline{U}:=\mathcal{C} (U\otimes I) \mathcal{C}^\dag$ is a logical $U$ operator where $\mathcal{C}$ is the encoding operator of  \Cref{sec:tableau}. 
There is no guarantee that $\overline{U}$ has a fault-tolerant implementation - for instance it may require a high-depth circuit.
In this work, we consider the following types of Clifford logical operators:
\begin{enumerate}
    \item \textbf{Single-qubit Transversal circuits}: Circuits with only single-qubit (Clifford) physical gates.
    \item \textbf{\textit{SWAP}-Transversal circuits}: Circuits with single-qubit Clifford gates and \textit{SWAP}s.
    \item \textbf{General Clifford circuits}: Circuits with single-qubit and general two-qubit physical gates.
\end{enumerate}

The first category involves transversal circuits and is considered fault-tolerant because an error in applying one of the single-qubit Cliffords does not spread errors to other qubits. 
In the case of  diagonal transversal logical operators (i.e. those composed of a series of single-qubit rotations around the $Z$-axis), efficient methods for finding these were set out in \cite{webster_transversal_2023}.
The methods in this paper identify transversal logical operators composed of other single-qubit Cliffords such as $H$ and products of $H$ and $S$ gates.

Depending on the qubit architecture, \textit{SWAP}-transversal circuits may also be considered fault-tolerant.
In some architectures, \textit{SWAP} gates can be accomplished without any direct interaction between the qubits and no physical operation involving both qubits simultaneously, and hence any faults in the process lead to independent errors on the qubits. 
Examples of such architectures include ion trap shuttle systems \cite{schoenberger_shuttling} and atom arrays \cite{evered_high-fidelity_2023}. 
In fixed layout architectures (e.g. superconducting), \textit{SWAP} gates can be made fault-tolerant via the use of auxiliary qubits, which may have their states reset during the process \cite{bravyi_high-threshold_2024}.\par

If \textit{SWAP} gates are viewed as a primitive instruction of our system, then similarly to \cite{moussa_transversal_2016}, the circuits with transversal operations and qubit permutations are fault-tolerant as the second category collapses to the first. This can happen for example on Rydberg atom\cite{xiao_effective_2024}, ion trap \cite{kielpinski_architecture_2002} or electron \cite{baart_single-spin_2016} qubit architectures. 

Hence, the fault tolerance of operations consisting of single-qubit Clifford gates and \textit{SWAP}s can vary depending on the architecture of the system, but can be viewed as a special case between transversal and general circuits.

The third category includes general Clifford circuits which can be of high depth and there is no guarantee that these can be implemented in a fault-tolerant way. 
Many qubit architectures have connectivity constraints which limit which qubits can be acted upon by multi-qubit gates.
The embedded code method outlined in \Cref{sec:embedded} allows for the search to be limited to gates which can be implemented within the connectivity constraints of the given device, though the circuits may still be of high depth. We restrict our search in relevant examples to circuits with few \textit{CNOT} and \textit{CZ} gates to exclude trivial circuit implementations such as decoding, applying a unitary and then re-encoding.

These circuits are fault-tolerant when no qubit within a code block is involved in more than one two-qubit gate. Low-depth circuits can potentially be implemented if they are pieceably fault-tolerant. This concept was introduced in \cite{yoder_universal_2016} and used in \cite{Quintavalle2023partitioningqubits} where circuits are partitioned and interleaved with error correction rounds of a possibly different code to ensure that low-weight physical errors are corrected before propagating to logical errors. Furthermore, logical gates where the physical entangling $2$ qubit gates (ie \textit{CNOT, CZ} or their generalizations) can be fault tolerant but with a reduced circuit distance. Specifically, if the distance of the code is $d$ then $r$ entangling $2$ qubit gates may reduce the circuit distance to $d-2r$ \cite{Manes2025distancepreserving}. Note that the additional resources and depth of these circuits can become quite large in this case. We only find automorphisms corresponding to physical circuits with entangling $2$ qubit gates using the embedded codes technique from  \Cref{sec:embedded}.

\subsection{Previous work}\label{sec:previous_work}
In \cite{calderbank1997}, the authors described a way to map quantum stabilizer codes to additive codes over $GF(4)$. Automorphisms over $GF(4)$, which are defined as permutations and multiplications by a field element are then used to define equivalent codes. To facilitate the automorphism group calculation, they provide a mapping of Pauli strings of length $n$ corresponding to the stabilizer generators to binary strings of length $3n$.
A method for calculating automorphism groups of linear codes over finite fields was developed in \cite{leon_computing_1982}. 
Here we aim to bring some intuition into why binary strings of length $3n$ are useful to represent Pauli operators by drawing parallels to the symplectic representation of stabilizer codes. 
We set out different mappings of Pauli operators to binary strings of length $2n$ which allow us to find specific types of circuits as code automorphisms.   
We also address issues such as Pauli corrections required to preserve the stabilizer group in circuits, since the original $GF(4)$ representation in \cite{calderbank1997} only works up to a sign of $\pm 1$. 

In \cite{grassl_leveraging_2013}, automorphism groups of codes are considered to be qubit permutations which preserve the codespace. 
The authors consider qubit permutations which map $X$-checks to $X$-checks and  $Z$-checks to $Z$-checks.
They then show that a larger set of fault-tolerant gates can be derived by using two copies of a CSS code and transversal \textit{CNOT} operators between the two blocks.
Our method accommodates  non-CSS stabilizer codes, and arbitrary combinations of physical \textit{CNOT}, $H$, $S$ and \textit{SWAP} gates.
In \cite{cross_universality}, the authors consider automorphism gates as a generalization of transversal logical operators. 
They show that a universal gate set cannot be generated using transversal or \textit{SWAP}-transversal operators on a code.
Furthermore, a condition is given for a particular logical action to be realised on a code via code automorphisms.

Recently, automorphisms have been used to find logical operators on quantum codes based on \textit{ZX}-dualities in which the $X$-checks of a CSS code are mapped to the $Z$-checks by combining qubit permutations and transversal $H$ gates. In \cite{moussa_transversal_2016}
\textit{ZX}-dualities were used to implement fault-tolerant logical gates on surface codes. 
The method was extended to qLDPC CSS codes in \cite{breuckmann_fold-transversal_2022, Quintavalle2023partitioningqubits} with the aim of implementing $H$ logical operators made from physical $H$ and \textit{SWAP} gates. They also give a construction based on such a \textit{ZX}-duality to perform a phase type logical operator made from physical $S, S^{\dag}$ and \textit{CZ} gates. Notably, the \textit{ZX}-duality method was referenced in \cite{bravyi_high-threshold_2024} and \cite{da_silva_demonstration_2024} as a way to implement $H$ gates on bivariate bicycle codes and the carbon code. Here we extend the method to non-CSS codes and show how to find a broader group of fault-tolerant logical operators composed of physical qubit swaps and $H$ gates.
We refer to these as \textit{SWAP}+$H$ gates. 
We generalize the method to \textit{SWAP}+$S$  and \textit{SWAP}+$\sqrt{X}$ gates and connect it to the general logical Clifford gates through automorphisms method.\par

\section{Automorphisms of Stabilizer Codes}\label{sec:auts_of_stab_codes}



In this Section we show how to find \textit{SWAP}-transversal logical operators of a stabilizer code by mapping the stabilizer generators to a binary linear code, finding the permutation automorphisms of the binary linear code, then mapping these back to valid logical Clifford operators.
We start this Section by using the two-block symplectic representation of the stabilizer generators to find to logical operators composed of $H$ and \textit{SWAP} gates. 
This generalizes the \textit{ZX}-dualities of \cite{breuckmann_fold-transversal_2022,Quintavalle2023partitioningqubits} to non-CSS codes.
We then show how to generalize the two-block method to find logical operators composed of either $S$ and \textit{SWAP} gates or $\sqrt{X}$ and \textit{SWAP} gates.
Finally, we introduce a $3$-block representation which allows us to find logical operators composed of single-qubit Cliffords in the group $\braket{H,S}$ and \textit{SWAP} gates.

\subsection{Logical Operators Composed of Physical \textit{SWAP} and Hadamard Gates}\label{sec:zx_duality}
We now show how to find all logical operators which can be implemented using physical \textit{SWAP} and $H$ gates. 
To do this, we use the two-block symplectic binary representation of Pauli operators outlined in \Cref{eq:standard_form}:
\begin{align}
G:=\left[
\begin{array}{c|c}
G_X & G_Z \\
\end{array}\label{hxz}
\right].
\end{align}

We then calculate the permutation automorphisms of the binary linear code with generator matrix $G$ whose codewords are $\braket{G}$, the span of the rows of $G$ over $\mathbb{F}_2$.
The \textbf{automorphism group} $\text{Aut}(\braket{G})$ of the binary linear code with generator matrix $G$ is the set of permutations of the bits of $G$ which map each codeword in $\braket{G}$ to a potentially different codeword (i.e. permutations of the bits which result in a permutation of the codewords);  we refer to these as \textbf{code automorphisms}.
A set of generators for $\text{Aut}(\braket{G})$ can be found by using the MAGMA computational algebra package.

In some cases, code automorphisms do not correspond to valid Clifford circuits constructed from $H$ and \textit{SWAP} gates acting on the stabilizer code.
To address this, we consider the intersection of  $\text{Aut}(\braket{G})$ with the automorphism group of the binary linear code $\braket{B}$ whose generator matrix is: 
\[
B:= \left[
\begin{array}{c|c}
I_n & I_n \\
\end{array}
\right].
\] 
The automorphism group of $\langle{B}\rangle$ is generated by the following elements:
\begin{enumerate}
    \item \textbf{$H$-type:} $(i,n+i)$. Exchanges the $X$ and the $Z$ block of the classical linear code. This corresponds to a physical $H$ gate on qubit $i$ of the stabilizer code.
    \item \textbf{\textit{SWAP}-type:} $(i,j)(n+i,n+j), i,j\leq n$. Exchanges the $X$ and the $Z$ components of indices $i$ and $j$ on the classical linear code. This corresponds to a physical \textit{SWAP} gate between qubits $i$ and  $j$ on the stabilizer code.  
\end{enumerate}
By restricting to automorphisms of $\braket{B}$, we ensure that code automorphisms correspond to a combination of $H$ and \textit{SWAP} gates acting on the stabilizer code.
We give more details on the derivation of this result in  \Cref{sec:aut_grp_intersection}.
\begin{example}\label{eg:513_H+SWAP}
We will now calculate the $H$+\textit{SWAP} circuits corresponding to automorphisms of the perfect $[[5,1,3]]$ code. We found the standard form of its stabilizer generators and logical basis in  \Cref{eq:n5code_sform,eq:n5code_sform_logicals} respectively. 
We thus find the automorphism group of the linear code on $10$ bits $\braket{G}$ with the above generator matrix and intersect it with the automorphism group of the binary linear code on $10$ bits $\braket{B}$ with generator matrix given by:
\[
B:=\left[
\begin{array}{c|c}
I_5 & I_5 \\
\end{array}
\right].
\]

The group $\text{Aut}(\braket{G}) \cap \text{Aut}(\braket{B})$ is of order $20$ and is generated by $3$ permutations. 
Two of these are qubit swaps which act as logical identities. 
There is also a non-trivial logical operator consisting of $H$ and \textit{SWAP}s which is shown in \Cref{fig:n5_hgate}, this is similar to the circuit previously found in \cite{yoder_universal_2016}.
This operator has the following action on the logical operators of the code:
\begin{align*}
    \bar{X}&=ZIIZX\mapsto XIZIX=-g_1 \bar{Z}\\
    \bar{Z}&=ZZZZZ\mapsto XXXXX=-g_1 g_2 g_3 g_4 \bar{X}.
\end{align*}
As $\bar{X}$ and $\bar{Z}$ map to $-\bar{Z}$ and $-\bar{X}$ respectively, this operator corresponds to a logical $\bar{Y}\bar{H}$.
We will describe how to find the logical action and adjust for the signs of mapped logical operators in more detail in \Cref{sec:pauli_correction}. \par
\begin{figure}[H]
    \centering
    \includegraphics[width=0.6\linewidth]{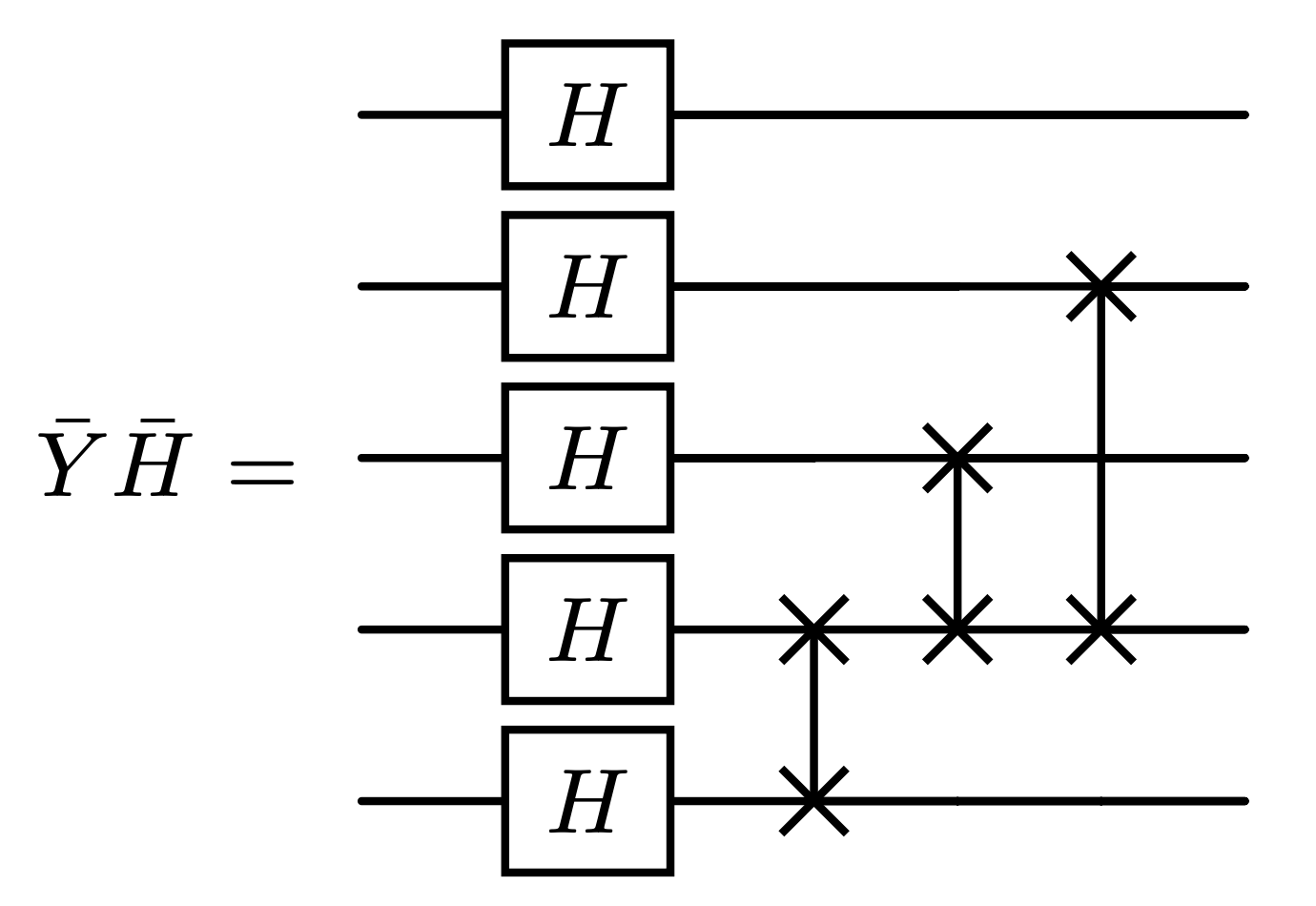}
    \caption{\textit{H+SWAP} gate of the [[5,1,3]] code.}
    \label{fig:n5_hgate}
\end{figure}
\end{example}

In \cite{breuckmann_fold-transversal_2022}, the authors introduce the concept of $ZX$-dualities for a class of CSS codes.  
$ZX$-dualities map $X$-checks of the code to $Z$-checks and vice versa and are implemented using $H$ and \textit{SWAP} operations.
The above method generalizes the concept of a $ZX$-duality to non-CSS codes and finds all Clifford logical operators which can be formed from \textit{SWAP}s and $H$s, if such operators exist.\par 

Furthermore, we can use alternative binary representations to find Clifford logical operators formed from \textit{SWAP} and $S\equiv\sqrt{Z}$ gates.
We do this by choosing the representation $[G_Z|G_X\oplus G_Z]$, where by $\oplus$ we denote the vector sum modulo $2$.
In this representation, swapping column $i$ in the $G_Z$-block with column $n+i$ in the $G_X\oplus G_Z$ block corresponds to applying an $S_i$ gate.
Similarly, we use the representation $[G_X|G_X \oplus G_Z]$ to find logical operators formed from \textit{SWAP} and $\sqrt{X}$ gates.

\subsection{Logical operators formed from physical single-qubit Clifford and \textit{SWAP} gates}\label{sec:swap_transversal}
We have so far shown how to find logical operators formed from \textit{SWAP}s and one of the single-qubit Cliffords. We now extend this to arbitrary combinations of single-qubit Cliffords and \textit{SWAP}s. This is done by extending the symplectic representation to a $3$-block representation as follows.

Starting from the symplectic representation of the stabilizer generators ${G}:=\left[\begin{array}{c|c}G_X & G_Z\end{array}\right]$, 
we append an all-zero block with $n$ columns and apply the binary transformation given by $$\mathcal{E} := \begin{bmatrix}I&0&I\\0&I&I\\I&I&I\end{bmatrix}$$ 
so that:
\begin{gather}
G_{\mathcal{E}}:=\left[\begin{array}{c|c|c}G_X&G_Z&0\end{array}\right]\mathcal{E}=
\notag\left[\begin{array}{c|c|c}G_X&G_Z&G_X\oplus G_Z
\end{array}\right]\label{sec:3bit}
\end{gather}
\begin{example}\label{eg:513_3-block}
    For the $[[5,1,3]]$ code of Example \ref{eg:513_standard_form}, the  $3$-block binary representation of the stabilizer generators is:
\begin{align*}
\begin{array}{|c|c|c|}
\hline
G_X & G_Z& G_X\oplus G_Z \\
\hline
10010&01100&11110\\
01001&00110&01111\\
10100&00011&10111\\
01010&10001&11011\\
\hline
\end{array}\label{513bit}
\end{align*}

\end{example}

To ensure that code automorphisms correspond to Clifford circuits on the stabilizer code, we take the intersection with the automorphism group of the (classical) linear code whose generator matrix is:
\[
B:=\left[
\begin{array}{c|c|c}
I_n & I_n & I_n \\
\end{array}
\right].
\]

As we can see, for this matrix, column swaps can correspond to different actions depending on their indices. Below we will use the cycle notation for permutations $(i,j,k)$ corresponding to the permutation that sends $i\mapsto j \mapsto k \mapsto i$. As before, $n$ is the number of physical qubits and let $i$ be some integer such that $1\leq i \leq n$. The indices of columns corresponding to qubit $i$ are $i,n+i,2n+i$, and the permutations correspond to the following gates: 
\begin{itemize}
    \item\textbf{$H$ type:} $(i,n+i)$  exchanges the $X$ and the $Z$ part of the stabilizer and hence corresponds to a physical $H$ gate on qubit $i$.
    \item\textbf{$S$ type:} $(n+i,2n+i)$  exchanges the $Z$ and the $XZ$ part of the stabilizer and corresponds to a physical $\sqrt{Z}$ gate (which is $S$ up to Pauli correction based on its phase as we will see in \Cref{sec:pauli_correction}).
    \item\textbf{\textit{SWAP} gates:} $(i,j)(n+i,n+j)(2n+i,2n+j)$ for $1\leq j \leq n, j\neq i$ corresponds to physical \textit{SWAP} gate between qubits $i$ and $j$. 
\end{itemize}
We provide a more extensive description of this mapping in  \Cref{sec:aut_grp_intersection} and show how to find the corresponding symplectic matrix representation in  \Cref{sec:perm2sym}.

Each automorphism, expressed as a column permutation matrix $P$, is associated with a symplectic matrix $\overline{U}_S$ which we calculate as follows. 
By examining each of the allowed gates, we find that for $\overline{U}_S$  a binary symplectic matrix and $W$  an invertible binary matrix which acts only on the $XZ$ block of the stabilizer generators:
\begin{align}
    \mathcal{E}P\mathcal{E}^{-1} = U_S \oplus W.
\end{align}
For simplicity, we identify the symplectic matrix $\overline{U}_S$ with the corresponding Clifford operator going forward. 
To ensure  $\overline{U}_S$ is a valid logical operator on the code, we may need to apply a Pauli correction and this is described in  \Cref{sec:pauli_correction}.

In  \Cref{tab:aut_family} we highlight the different automorphism finding algorithms based on what physical circuits they allow us to find.
\renewcommand{\arraystretch}{1.2}
\begin{table}[h]
\centering
\begingroup
\medmuskip=0.1mu
\thinmuskip=0.1mu
$$\begin{array}{|l|l|l|l|}
\hline
\textbf{Physical} &  \textbf{Binary}  &  \textbf{Transf}&  \textbf{Intersect}\\
\textbf{Gate Types} &  \textbf{Representation}  &  \textbf{Matrix } \mathcal{E}&  \textbf{Code } B\\
\hline
\text{1. } H + \textit{SWAP} & [G_X|G_Z] & \begin{bmatrix}I&0\\0&I\end{bmatrix} & \begin{bmatrix}I&I\end{bmatrix}\\
\hline
\text{2. } S + \textit{SWAP} & [G_Z|G_X\oplus G_Z] & \begin{bmatrix}0&I\\I&I\end{bmatrix} & \begin{bmatrix}I&I\end{bmatrix}\\
\hline
\text{3. } \sqrt{X} + \textit{SWAP} & [G_X|G_X\oplus G_Z] & \begin{bmatrix}I&I\\0&I\end{bmatrix} & \begin{bmatrix}I&I\end{bmatrix}\\
\hline
\text{4. } \braket{H,S} + \textit{SWAP} & [G_X|G_Z|G_X\oplus G_Z] & \begin{bmatrix}I&0&I\\0&I&I\\I&I&I\end{bmatrix} & \begin{bmatrix}I&I&I\end{bmatrix}\\
\hline
\end{array}$$
\endgroup
\caption{Family of automorphism algorithms: we show how to find fault-tolerant logical operators formed from \textit{SWAP} and various single-qubit Clifford gate sets. For a stabilizer code with check matrix $G :=[G_X|G_Z]$, we choose a transformation matrix $\mathcal{E}$ to create a generator matrix $G_\mathcal{E}$ for a related binary linear code. 
We then find the automorphism group $\text{Aut}(\braket{G_\mathcal{E}})$ using a computational algebra system such as MAGMA. Restricting the automorphism group to $\text{Aut}(\braket{G_\mathcal{E}}) \cap \text{Aut}(\braket{B})$  ensures that the logical operators use only the desired gate set.}\label{tab:aut_family}
\end{table}

\section{Pauli Correction to Ensure Signs of stabilizer Group are Preserved}\label{sec:pauli_correction}
In this section we show how to apply a Pauli correction to ensure that the Clifford operators $\bar{U}_S$ of  \Cref{sec:swap_transversal} identified via code automorphisms preserve the phases of the stabilizer generators. 
This step is necessary because symplectic matrices correspond to Clifford operators modulo Pauli operators and phases.

The Clifford operator $\overline{U}_S$ preserves the stabilizer group, but only up to a sign of $\pm 1$. 
This is because for a non-trivial Pauli operator $A$ to be a valid stabilizer, it must have eigenvalues of $\pm 1$ and so $A^2 = I$. 
Note that $-A$ is also a valid stabilizer with eigenvalues $\pm 1$, but that $\pm iA$ is not a valid stabilizer as it has eigenvalues $\pm i$.
Conjugation by a Clifford operator $U$ results in an operator which squares to $I$ because:
\begin{align}
    (UAU^\dag)^2 = UAU^\dag UAU^\dag= UA^2U^\dag = I.
\end{align}
Accordingly, the Clifford operator associated with $\overline{U}_S$ maps stabilizer generators to elements of the stabilizer group up to a sign of $\pm 1$ but never $\pm i$.

For each stabilizer generator $G_i$ with destabilizer $R_i$, we perform the following steps.
We first determine the phase applied to each stabilizer generator by $\bar{U}_S$ so that for some binary vectors $\mathbf{x,z}$ and integer $u$ modulo $4$:
\begin{align}
    \bar{U}_\text{S}G_i \bar{U}_\text{S}^\dag = i^u X(\mathbf{x})Z(\mathbf{z}).\label{eq:conj_phase}
\end{align}
If $\bar{U}_S$ is a valid logical operator, each $G_i$ should be mapped to a product of stabilizer generators. 
As a result, the binary vector $(\mathbf{x|z})$ is a linear combination of the rows of the check matrix $G$ modulo $2$ and for some binary vector $\mathbf{g}$ of length $n-k$, we have:
\begin{align}
    (\mathbf{x|z}) = \mathbf{g}G.
\end{align}
Determining $\mathbf{g}$ allows us write the mapped operator as a product of the stabilizer generators so that for some integer $v$ modulo $4$:
\begin{align}
    \prod_{0 \le j < n-k} G_j^{\mathbf{g}[j]} = i^v X(\mathbf{x})Z(\mathbf{z}).\label{eq:stabilizer_product}
\end{align}
If the phase applied by $\bar{U}_S$ is different to the phase of the product of stabilizer generators, we apply the destabilizer $R_i$ to the Pauli correction.
We now examine these steps in more detail.

\subsection{Phase Applied by a Clifford Circuit}\label{sec:circuit_phase}
The logical operators found by the methods of \Cref{sec:swap_transversal} are \textit{SWAP}-transversal circuits.
As \textit{SWAP} gates do not apply any phase, the phase applied by the circuit $\bar{U}_S$ is determined by the transversal single-qubit Clifford gates.
Single-qubit Clifford gates can apply a phase of $-1$ in the case of the following Pauli operators:
\begin{align}
    HYH &= -Y;\\
    SYS^\dag &= -X;\\
    \sqrt{X}Z\sqrt{X}^\dag &= -Y.
\end{align}
We capture the phase applied by single-qubit Cliffords by pre-pending an integer modulo $4$ to the $3$-block binary representation (\Cref{sec:3bit}) of single-qubit Paulis so that $(p, a, b, c):= (p, x, z, x\oplus z)$.
There are six single-qubit Clifford operators (up to Pauli factors and phases) and the update rules for their action on the single-qubit Pauli operators are set out in  \Cref{tab:clifford_actions}. 
The update rules allow us to calculate the phase $p$ of $\bar{U}_S G_i \bar{U}_S^\dag$ in  \Cref{eq:conj_phase} by computing a dot product on the 3-block representation.
\renewcommand{\arraystretch}{1.2}
\begin{table}[H]
\centering
\begingroup
\medmuskip=0.1mu
\thinmuskip=0.1mu
$$\begin{array}{|c|c|c|c|l|}
\hline
\textbf{Clifford} &  \mathbf{X}   &\mathbf{Y}  &\mathbf{Z} &  \textbf{Action on }\\
\textbf{Gate}& (0,1,0,1) & (1,1,1,0)&(0,0,1,1)&(p,a,b,c)\\
\hline
I & X&Y&Z& (p,a,b,c) \\
H & Z & -Y &X & (p+c-a-b,b,a,c) \\
S:=\sqrt{Z}&Y&-X&Z&(p+a,a,c,b)\\
\sqrt{X}:=HSH&X&Z&-Y&(p-b,c,b,a)\\
\Gamma:=HS^\dag&Y&Z&X&(p+c-b,c,a,b)\\
\Gamma^\dag:=SH &Z&X&Y&(p+c-a,b,c,a)\\
\hline
\end{array}$$
\endgroup
\caption{Single-qubit Clifford gates and action on\\ phase + $3$-block representation.}\label{tab:clifford_actions}
\end{table}

\begin{example}\label{eg:gamma}
The action of $\Gamma:=HS^\dag$ on single-qubit Paulis is:\\
$\Gamma(X) = \Gamma(0,1,0,1) = (1,1,1,0) = iXZ=Y$\\
$\Gamma(Y) = \Gamma(1,1,1,0) = (1-1,0,1,1) = Z$\\
$\Gamma(Z) = \Gamma(0,0,1,1) = (1-1,1,0,1) = X$.
\end{example}
\subsection{Phase of Mapped Stabilizer}\label{sec:stabilizer_product}
A valid logical operator maps each stabilizer generator to an element of the stabilizer group and this may be a product of various different stabilizer generators.
We can express any Pauli operator as a product of stabilizer, destabilizer and logical generators by using the tableau representation $\tau$ for the stabilizer code as set out in  \Cref{sec:tableau}.
For a stabilizer code with stabilizer, destabilizer, \textit{X}- and \textit{Z}-logical generators in binary form $G,R,L_X$ and $L_Z$ respectively, the tableau $\tau$ is a symplectic matrix of form:
\begin{align}
\tau := 
\left[\begin{array}{c}
G\\ L_X\\\hline R\\L_Z
\end{array}\right].
\end{align}
As $\tau$ is a symplectic matrix, its inverse is given by $\Omega \tau^T \Omega$ because:
\begin{align}
    \tau \Omega \tau^T \Omega = \Omega^2 = I.
\end{align}
Hence an unsigned Pauli operator $X(\mathbf{x})Z(\mathbf{z})$ can be expressed as a product of stabilizer, destabilizer and logical Pauli generators by calculating the vector $\mathbf{b}$ as follows:
\begin{align}
    \mathbf{b}&:= (\mathbf{x}|\mathbf{z})\Omega \tau^T \Omega\\
    &:=(\mathbf{g}|\mathbf{a}_X|\mathbf{r}|\mathbf{a}_Z);\label{eq:b_vector}
\end{align}
where $\mathbf{g}$ and $\mathbf{r}$ are binary vectors of length $n-k$ and $\mathbf{a}_X$ and $\mathbf{a}_Z$ are of length $k$ so that:
\begin{align}
    (\mathbf{x|z})&= \mathbf{b}\tau = \mathbf{g}G + \mathbf{a}_X L_X + \mathbf{r}R + \mathbf{a}_Z L_Z.
\end{align}
As we require that $(\mathbf{x|z})$ is in the span of the stabilizer generators $G$, the vectors $\mathbf{a}_X,\mathbf{a}_Z$ and $\mathbf{r}$ should be zero so that $(\mathbf{x|z}) = \mathbf{g}G$.
The phase $v$ of  \Cref{eq:stabilizer_product} results from the product of stabilizers specified by the non-zero entries of $\mathbf{g}$.

\subsection{Application of Destabilizer}\label{sec:destabilizer_action}
We now show applying a destabilizer has the desired effect of updating signs for mapped stabilizer generators.
Assume that $u\ne v$ in  \Cref{eq:conj_phase,eq:stabilizer_product} so that:
\begin{align}
     \bar{U}_\text{S}G_i \bar{U}_\text{S}^\dag = - \prod_{0 \le j < n-k} G_j^{\mathbf{g}[j]}.
\end{align}
If we instead conjugate $G_i$ by $\bar{U}_\text{S}R_i$, where $R_i$ is the destabilizer of $G_i$, we have:
\begin{align}
     \bar{U}_\text{S}R_iG_i R_i\bar{U}_\text{S}^\dag &=\bar{U}_\text{S}(R_iG_i R_i)\bar{U}_\text{S}^\dag \\&= - \bar{U}_\text{S}G_i \bar{U}_\text{S}^\dag \\
     &= \prod_{0 \le j < n-k} G_j^{\mathbf{g}[j]}.
\end{align}
Because $R_i$ commutes with all other stabilizer generators, the sign of each mapped stabilizer generator can be updated independently.
Accordingly, we choose the Pauli correction $\bar{U}_P$ which is the product of the destabilizers $R_i$ where the corresponding mapped stabilizer generator is an element of the stabilizer group up to a sign of $-1$.

\section{Determining Logical action of a Clifford circuit}\label{sec:logical_action}

In this section, we show how to determine the logical action of a Clifford logical operator $\bar{U}_S$.
We will use the tableau form of  \Cref{sec:tableau} 
to apply a Pauli correction to $\bar{U}_S$ and to describe the logical action as a $2k\times 2k$ symplectic matrix.

For each logical Pauli generator $L_i$, we determine the mapped operator $\bar{U}_\text{S}L_i \bar{U}_\text{S}^\dag = i^u X(\mathbf{x})Z(\mathbf{z})$ as described in  \Cref{sec:circuit_phase}.
We then calculate the vector  $\mathbf{b} =(\mathbf{g}|\mathbf{a}_X|\mathbf{r}|\mathbf{a}_Z)$ of  \Cref{eq:b_vector}.
The vectors $\mathbf{a}_X$ and $\mathbf{a}_Z$ describe the product of logical $X$ and $Z$ generators respectively which appear in the mapped logical operator.
Up to a Pauli operator, the logical action of $\bar{U}_S$ is given by the symplectic matrix $U_\text{ACT}$ whose $i$th row is given by the length $2k$ vector $(\mathbf{a}_X|\mathbf{a}_Z)$.
A logical $Y$ operator is represented as $XZ = -iY$, so we adjust the logical action by multiplying by the phase $i^{-\mathbf{a}_X \cdot \mathbf{a}_Z}$.
The \textbf{logical action phase adjustment} ensures that logical Pauli generators, which square to $I$, map to products of stabilizer and logical Pauli generators which also square to $I$.
For example, this step ensures that a logical $\bar{S}$ maps $\bar{X}$ to $\bar{Y} = i\bar{X}\bar{Z}$ rather than $\bar{X}\bar{Z}$, which squares to $-I$.
This also guarantees that the logical action is correct up to a sign of $\pm 1$.

We ensure that the sign of the logical action is correct by applying a Pauli correction using a similar method to that outlined in  \Cref{sec:pauli_correction}. 
This time, we use the vector $\mathbf{b}$ to calculate a product of stabilizer and logical generators, apply the logical action adjustment phase $i^{-\mathbf{a}_X \cdot \mathbf{a}_Z}$ and check whether this has the same phase as the mapped logical operator. 
If not, we apply the destabilizer of the logical operator  to update the sign. 
For a logical $\bar{X}_i$ corresponding to row $i$ of $L_X$, the destabilizer is the anti-commuting logical $\bar{Z}_i$ corresponding to row $i$ of $L_Z$.
As the logical generators commute with all other stabilizer and logical generators, this allows us to update the signs independently. 
A combined algorithm to calculate the Pauli correction $\bar{U}_P$ and the logical action $U_\text{ACT}$ is set out in  \Cref{alg:pauli_correction}.

\section{Representation of Symplectic Matrices as Circuits}\label{sec:sym2circuits}
Both the physical implementations of logical operators and their logical actions have so far been presented as a symplectic matrix plus a Pauli correction. 
In this section, we show how we translate this representation into a circuit composed of single- and 2-qubit Clifford gates.

In  \Cref{thm:circuit} in  \Cref{sec:clifford_decomp_proof} we show that any symplectic matrix can be written as $U = U_AU_BU_CU_H$ where:
\begin{itemize}
    \item $U_A$ represents a product of $C(X,X)$ and $\sqrt{X}$ operators;
    \item $U_B$ is a product of \textit{CZ} and $S$ operators;
    \item $U_C$ is a product of \textit{CNOT} operators; and
    \item $U_H$ is a product of single-qubit $H$ operators.
\end{itemize}
This leads to a natural decomposition of Clifford unitaries in $6$ layers of gates as $H \text{-} CNOT \text{-} S \text{-} CZ \text{-}\sqrt{X}\text{-}C(X,X)$.

\section{Multi-qubit Physical Gates via Automorphisms of Embedded Codes}\label{sec:embedded}
We have so far been working with \textit{SWAP}-transversal logical Clifford operators which consist of physical \textit{SWAP} and single-qubit Clifford operators.
The embedded code technique allows us to find a broader range of logical operators, including those which employ two-qubit \textit{CNOT} and \textit{CZ} physical gates.
The resulting logical Clifford operators may be high depth and are not guaranteed to be fault-tolerant.
The technique does allow for restrictions on which qubits can interact via two qubit gates and so may be useful for certain device architectures.

The embedded code technique was introduced in \cite{webster_transversal_2023} as a method for finding logical operators of CSS codes involving multi-qubit diagonal gates.
In the present work, we generalize the method to non-CSS codes and to non-diagonal Clifford gates.

This method aims to describe the action of physical two-qubit entangling gates, such as \textit{CNOT} and \textit{CZ} as an automorphism of a larger code. Specifically, the method works by embedding the original $n$ qubits to a code with $$n+\binom{n}{2}$$ qubits, where the extra $\binom{n}{2}$ qubits are initialized in the $|0\rangle$  state. We then make use of the following circuit identities:



\begin{figure}[H]
    \centering
    \includegraphics[width=\linewidth]{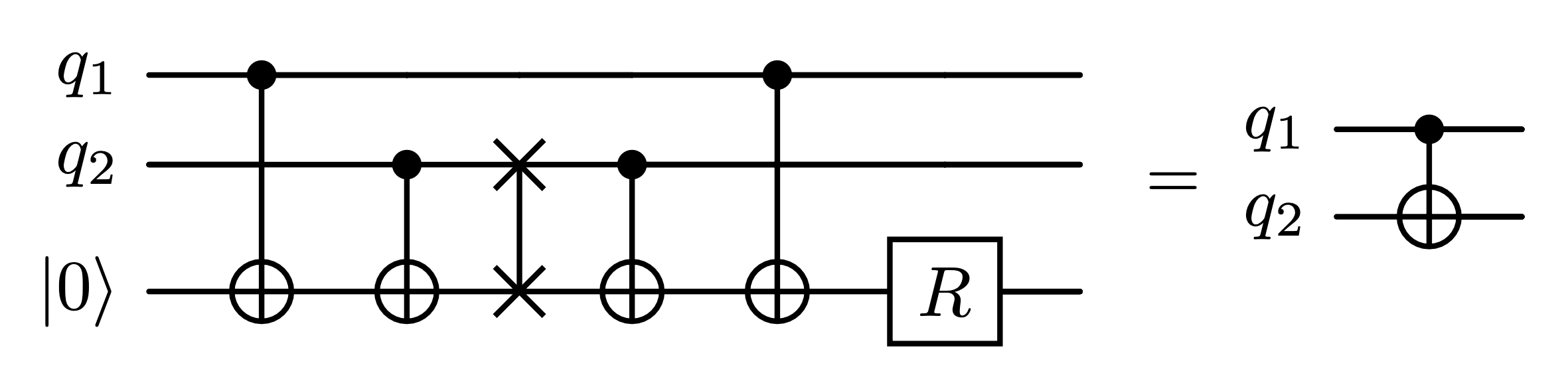}
\end{figure}
\begin{figure}[H]
    \centering
    \includegraphics[width=\linewidth]{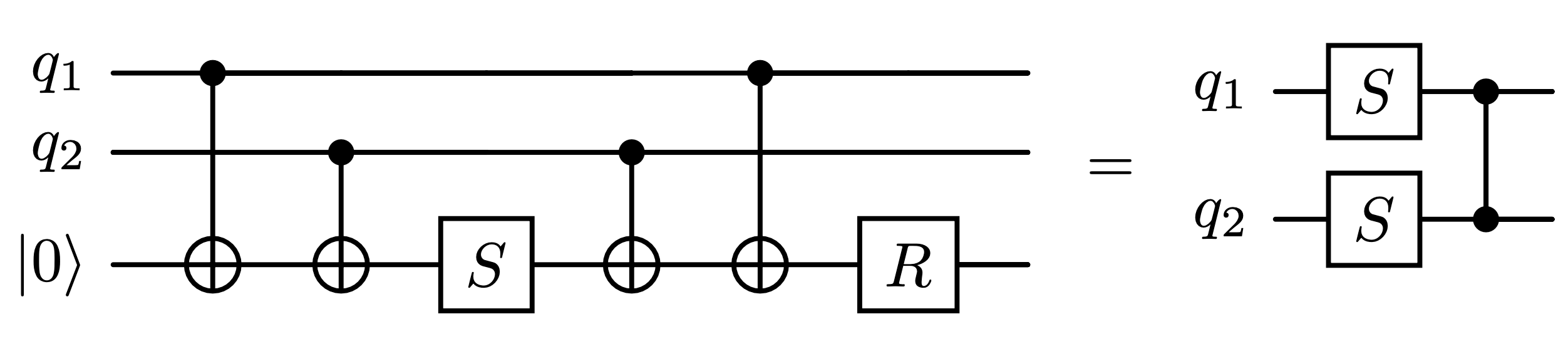}
\end{figure}

Where $R$ is the reset instruction. Since we don't actually use the extra qubits, the operation is equivalent to a \textit{CNOT} or \textit{CZ} gate.

We construct an embedded code by defining a $m \times n$ binary matrix $M$ whose rows are of weight two. 
Each row of $M$ corresponds to a pair of qubits in the original codespace.
Connectivity constraints of a particular device can be encoded by only including in $M$ rows where the corresponding two-qubit gates are available. We provide more details about the definition and calculation of the embedded code in \Cref{app:embedded_codes}.

For a stabilizer code with check matrix $G:=[G_X|G_Z]$, the check matrix of the embedded code on $n+m$ qubits is given by:
\begin{align}
   G_V:= \left[\begin{array}{cc|cc}
         G_X& G_X M^T & G_Z &0  \\
         0&0&M&I
    \end{array}\right]\label{eq:embedded_code}
\end{align}
In  \Cref{sec:embedded_code_auts}, we derive the relations between permutations of the columns of the check matrix of the embedded code and the corresponding single- and two-qubit Clifford operations on the original codespace. We include the results below in  \Cref{tab:embedactions}. 
In  \Cref{sec:422_gates} we show that we find a wider range of logical Clifford operators via the embedded code method versus the methods of  \Cref{sec:swap_transversal} for the $[[4,2,2]]$ code.
\renewcommand{\arraystretch}{1.2}
\begin{table}[H]
\centering
\begingroup
\medmuskip=0.1mu
\thinmuskip=0.1mu
$$\begin{array}{|c|c|}
\hline
\textbf{Embedded Codespace} &  \textbf{Original Codespace} \\
\hline
S_{(12)} & S_{1}S_{2}CZ_{12}\\
\hline
\textit{SWAP}_{(12),2} & \textit{CNOT}_{12}\\
\hline
\end{array}$$
\endgroup
\caption{Actions on Embedded Codespace and their corresponding physical gates on the original code. The notation $S_{(12)}$ indicates an $S$ operator applied to the auxiliary qubit represented by the pair of physical qubits $(1,2)$.}\label{tab:embedactions}
\end{table}

\section{Examples and results}\label{sec:examples_and_results}

\subsection{5-qubit perfect code}\label{sec:513_gates}
\begin{figure}[H]
    \centering
    \includegraphics[width=0.7\linewidth]{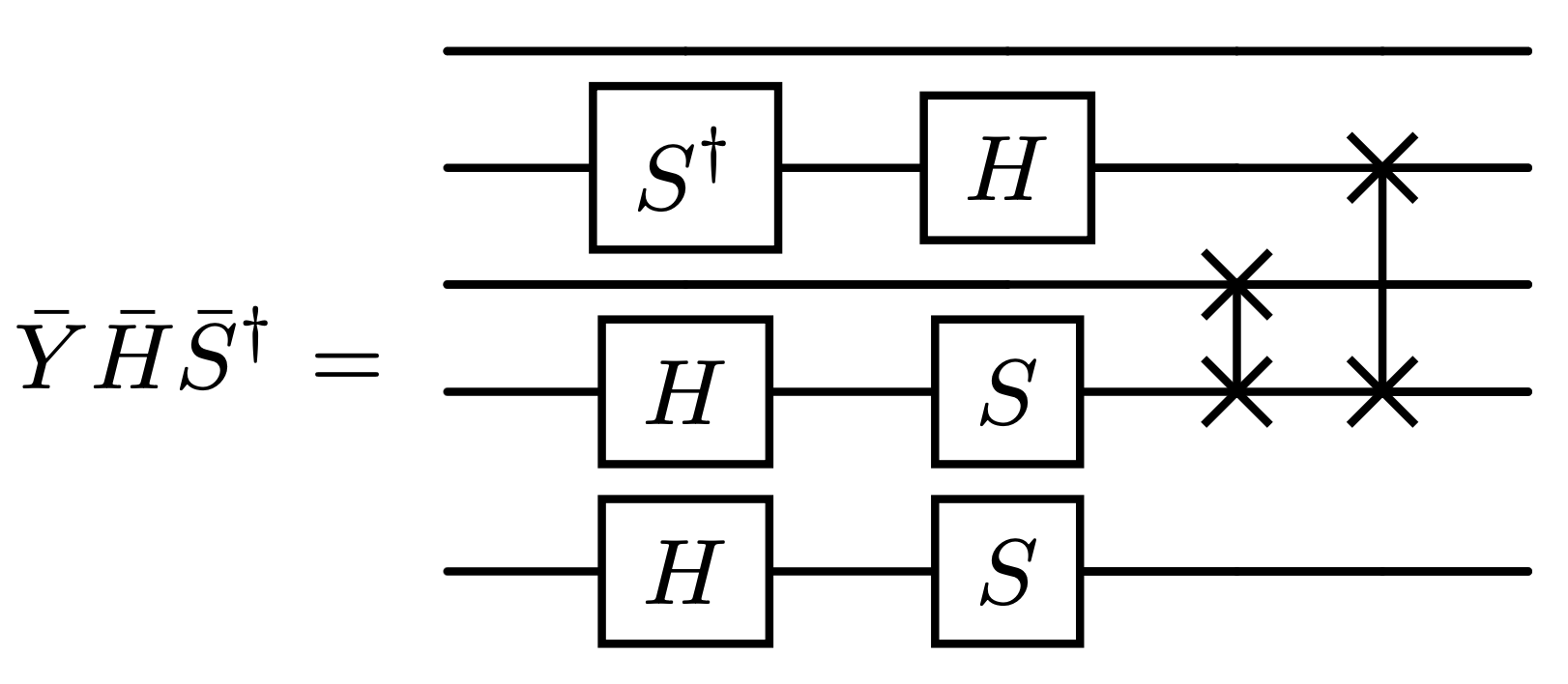}
    \caption{\textit{SWAP}-Transversal $\bar{\Gamma} = \bar{H} \bar{S}^\dagger$ gate of the [[5,1,3]] code with a $\bar{Y}$ correction.}
    \label{fig:n5k1_gamma}
\end{figure}

\begin{figure}[H]
    \centering
    \includegraphics[width=0.5\linewidth]{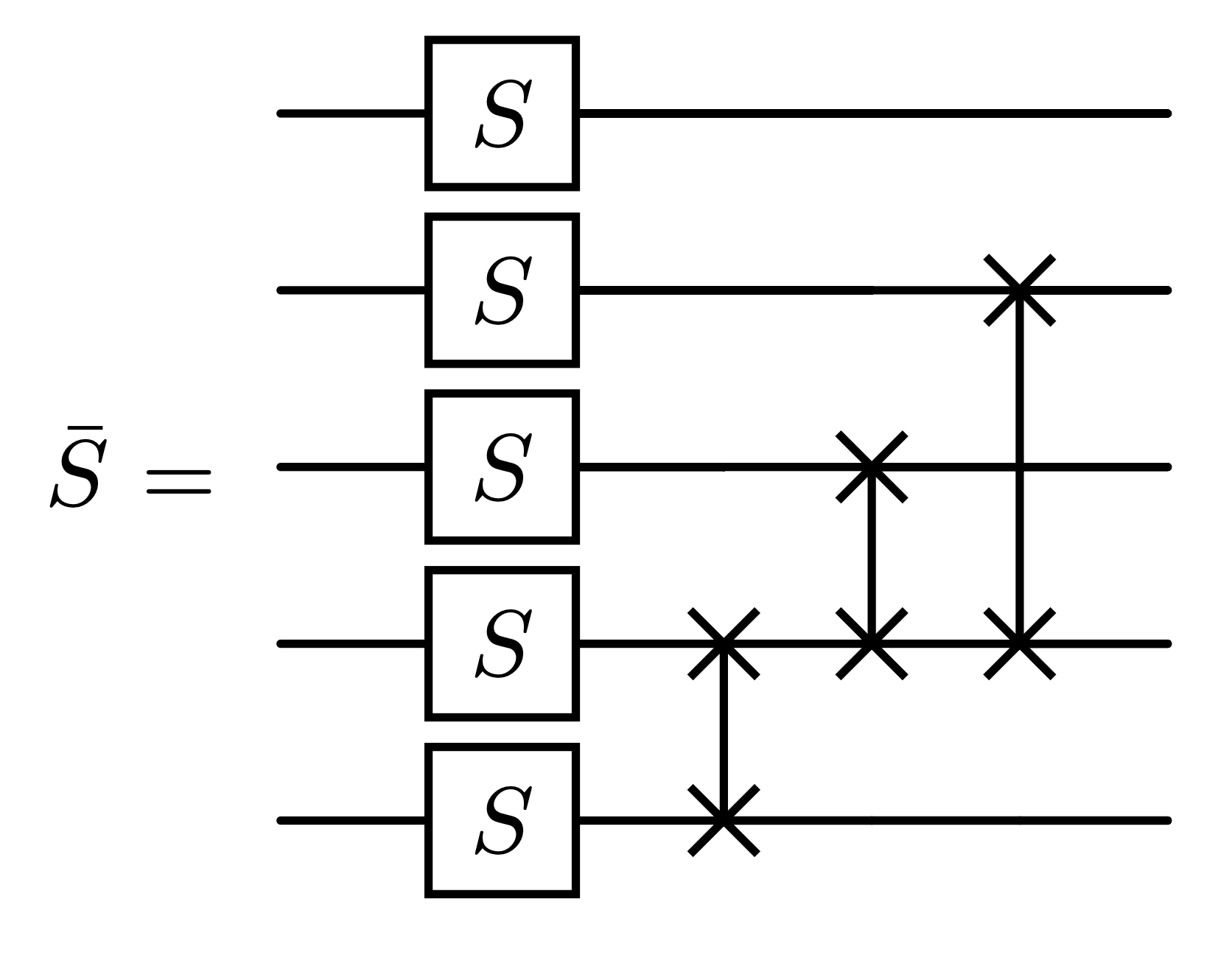}
    \caption{\textit{SWAP}-transversal $\bar{S}$ gate of the [[5,1,3]] code.}
    \label{fig:n5_sgate}
\end{figure}

We applied the algorithms of  \Cref{sec:auts_of_stab_codes,sec:pauli_correction} to the $5$-qubit perfect code with code parameters [[5,1,3]]. 
This code has a well-known transversal logical $\Gamma = HS^\dagger$ gate \cite{yoder_universal_2016}. Here we find a \textit{SWAP}-transversal circuit that implements the same logical gate up to a Pauli correction. The \textit{SWAP}-transversal alternative in \Cref{fig:n5k1_gamma} uses fewer single-qubit physical Clifford gates. \Cref{fig:n5_sgate} illustrates the \textit{SWAP}-transversal implementation of an $S$ gate. 

\noindent\href{https://github.com/hsayginel/autqec/tree/main/examples}{Our results are available as an interactive notebook.}

\subsection{4-qubit code}\label{sec:422_gates}
We now illustrate the 3-block method of  \Cref{sec:swap_transversal} by applying it to the $[[4,2,2]]$ code which has the following stabilizer generators and logical operators. 
$$\mathbf{G}:=\begin{bmatrix}
    XXXX\\
    ZZZZ
\end{bmatrix},\;\;\begin{bmatrix}
    L_X\\
    L_Z
\end{bmatrix}=\begin{bmatrix}
    XIIX\\
    XIXI\\
    \hline
    ZIZI\\
    ZIIZ\\
\end{bmatrix}$$
The resulting logical operators are shown in  \Cref{tab:n4k2d2}. 
Note that we do not obtain the full logical Clifford group because we do not have logical $H$ and $S$ addressing individual logical qubits.
\renewcommand{\arraystretch}{1.2}
\begin{table}[H]
    \centering
    \begin{tabular}{|c|c|c|}
        \hline
        \textbf{Gate} & \textbf{Circuit} & \textbf{Type}\\
        \hline
        $\overline{H}_1\overline{H}_2$ & $H_1H_2H_3H_4\textit{SWAP}_{3,4}$ & Transversal\\
        \hline
        $\overline{CZ}$ & $S_1^\dag S_2^\dag S_3 S_4$ & Transversal\\
        \hline
        $\overline{CNOT}_{1,2}$ & $\textit{SWAP}_{2,4}$ & \textit{SWAP}-transversal\\
        \hline
        $\overline{CNOT}_{2,1}$ & $\textit{SWAP}_{2,3}$& \textit{SWAP}-transversal\\
        \hline
        $\overline{\textit{SWAP}}_{1,2}$ & $\textit{SWAP}_{3,4}$ & \textit{SWAP}-transversal\\
        \hline
    \end{tabular}
    \caption{\textit{SWAP}-transversal Logical Clifford gates of the [[4,2,2]] code.}
    \label{tab:n4k2d2}
\end{table}
We do obtain a full set of logical Clifford operators for the $[[4,2,2]]$ code via the embedded code technique of  \Cref{sec:embedded}, though these are not \textit{SWAP}-transversal and so are not necessarily fault-tolerant.
We choose the embedding matrix $\begin{bmatrix}I\\M\end{bmatrix}$ where $M$ is the binary matrix whose rows are the binary vectors of length $4$ and weight $2$. We add $m:= \binom{4}{2} = 6$ auxiliary qubits and the check matrix of the embedded code is:
\begin{align*}
   G_V&:= \left[\begin{array}{cc|cc}
         G_X& G_X M^T & G_Z &0  \\
         \hline
         0&0&M&I
    \end{array}\right]\\
    &= \left[\begin{array}{cc|cc}
    1111&000000&0000&000000\\
    0000&000000&1111&000000\\
    \hline
    0000&000000&1100&100000\\
    0000&000000&1010&010000\\
    0000&000000&1001&001000\\
    0000&000000&0110&000100\\
    0000&000000&0101&000010\\
    0000&000000&0011&000001
\end{array}\right]
\end{align*}
We find circuits implementing $S$ and $\sqrt{X}$ on each logical qubit using  \Cref{alg:lo_search} which searches for implementations of logical operators with a desired action: 
\renewcommand{\arraystretch}{1.2}
\begin{table}[H]
    \centering
    \begin{tabular}{|c|c|c|}
        \hline
        \textbf{Gate} & \textbf{Circuit} & \textbf{Type}\\
        \hline
        $\overline{S}_1$ & $S_1 S_3 CZ_{1,3}$ & General Clifford\\
        \hline
        $\overline{S}_2$ & $S_1 S_4 CZ_{1,4}$ & General Clifford\\
        \hline
        $\overline{\sqrt{X}}_1$ & $\sqrt{X}_1 \sqrt{X}_4 C(X,X)_{1,4}$ & General Clifford\\
        \hline
        $\overline{\sqrt{X}}_2$ & $\sqrt{X}_1 \sqrt{X}_3 C(X,X)_{1,3}$ & General Clifford\\
        \hline
    \end{tabular}
    \caption{Logical Clifford gates of the [[4,2,2]] code via embedded code technique.}
\end{table}

\noindent\href{https://github.com/hsayginel/autqec/tree/main/examples}{Our results are available as an interactive notebook.}

\subsection{Best-Known-Distance Codes with $k=1$}\label{sec:codetable_gates}
We then applied the algorithms in  \Cref{sec:swap_transversal,sec:pauli_correction} to best-known-distance codes with $k=1$ logical qubit from \url{codetables.de} \cite{Grassl:codetables} on up to $n=30$ qubits. 
Our results are presented in  \Cref{tab:k1_codes}. 
We observe that most $k=1$ codes have the full single-qubit Clifford group, achieved via \textit{SWAP}-transversal gates. 
On the other hand, when restricting the logical gate implementation to only transversal operations, we can only achieve a small subset of these gates. 
In \Cref{tab:k1_trans} we list the transversal gate implementations.

\renewcommand{\arraystretch}{1.2} 
\begin{table}[H]
    \centering
    \begin{tabular}{|c|c|}
        \hline
        \textbf{Logical gates} & \textbf{n} \\
        \hline
        $\bar{I}$ & $ 4, 16, 28$ \\
        \hline
        $\bar{S}$ & $24$ \\
        \hline
        $\bar{S}\bar{H}$ & $10-15$\\
        \hline
        $\langle \bar{S}, \bar{H} \rangle$ & all other codes with $n\leq30$ \\
        \hline
    \end{tabular}
    \caption{\textit{SWAP}-transversal gates of $[[n,1,d]]$ codes up to $n\leq 30$ physical qubits.}
    \label{tab:k1_codes}
\end{table}

\renewcommand{\arraystretch}{1.2}
\begin{table}[H]
    \centering
    \begin{tabular}{|c|c|}
        \hline
        \textbf{Logical gates} & \textbf{n} \\
        \hline
        $\bar{I}$ & all other codes with $n\leq30$ \\
        \hline
        $\bar{S}$ & $2,3$ \\
        \hline
        $\bar{S}\bar{H}$ & $5,17,25,29$\\
        \hline
        $\langle \bar{S}, \bar{H} \rangle$ & $7$ \\
        \hline
    \end{tabular}
    \caption{Transversal gates of $[[n,1,d]]$ codes up to $n\leq 30$ physical qubits.}
    \label{tab:k1_trans}
\end{table}
\noindent \href{https://github.com/hsayginel/autqec/tree/main/codetables}
{Our results} for $k=1$ best-known distance codes indicate that small non-CSS codes, when combined with \textit{SWAP} gates, can realize logical $S$ and $H$ gates with lower physical qubit overhead than is possible using other codes, such as color codes, with comparable distance. These codes may thus offer practical advantages for quantum architectures with limited qubit resources.



\subsection{Bivariate Bicycle (BB) Codes}\label{sec:bb_code_gates}
The bivariate bicycle codes were introduced in \cite{bravyi_high-threshold_2024} and have generated much interest from the quantum error correction community.
 One of the main properties of these codes is that their logical qubits can be partitioned into two blocks of equal size, called the unprimed and primed blocks. The authors then identify two types of \textit{SWAP}-transversal gates:

\begin{enumerate}
    \item \textbf{Grid translation automorphisms:} \textit{SWAP}s along the Tanner graph which correspond to logical \textit{CNOT}s between qubits in the same logical block.

    \item \textbf{$ZX$-dualities:} logical global $H$s and \textit{SWAP}s between pairs of qubits across $2$ logical blocks.
\end{enumerate}

\begin{figure}[H]
    \centering
        \includegraphics[width=\linewidth]{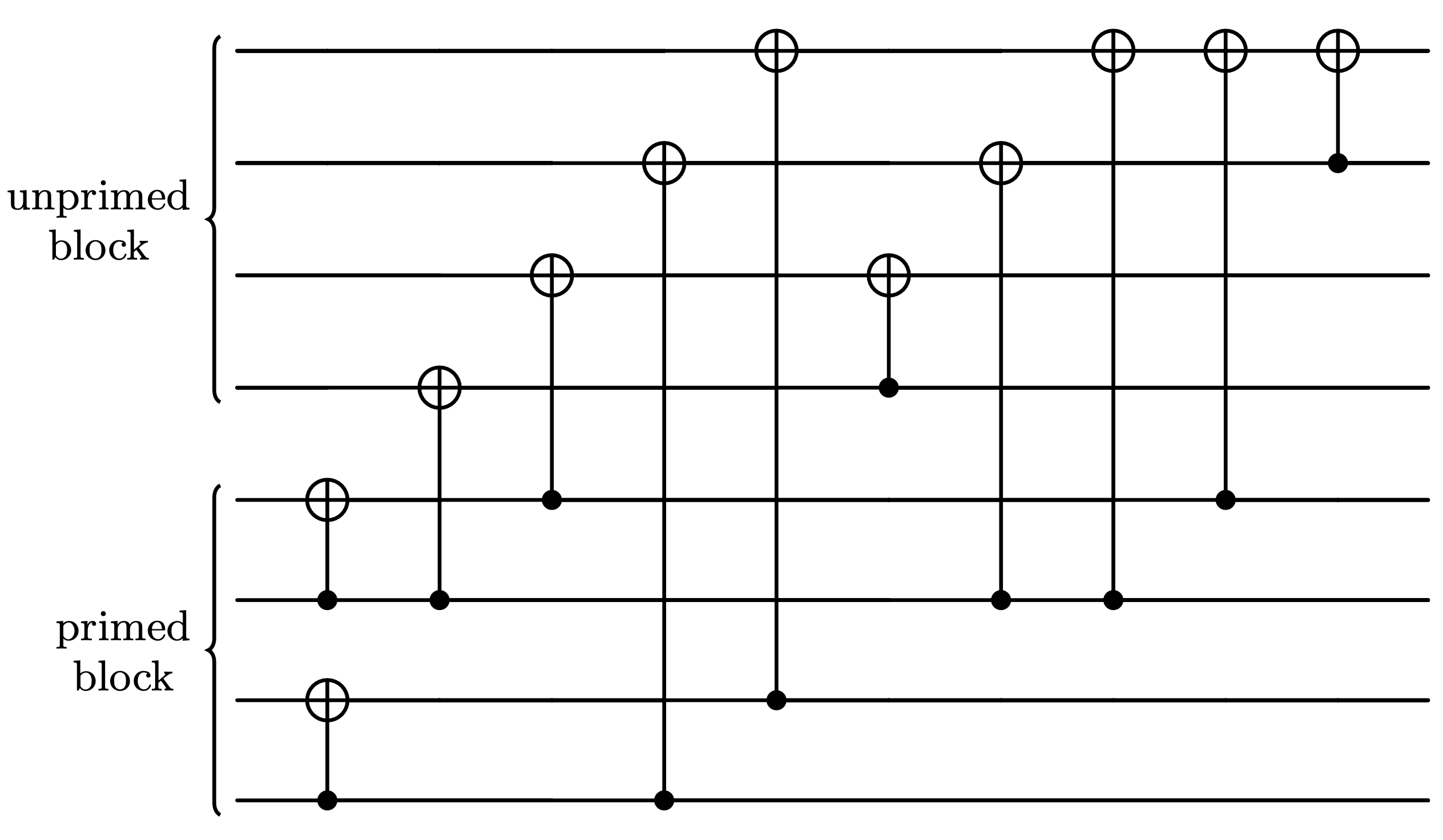}
    \caption{Logical action of the additional automorphism group generator for the $[[108,8,10]]$ Bivariate Bicycle Code as a \textit{CNOT} circuit.}
    \label{fig:n108k8d10}
\end{figure}

Using the method of  \Cref{sec:zx_duality}, found $H$+\textit{SWAP} logical operators for the codes of \cite{bravyi_high-threshold_2024} for up to $n=360$ qubits. This allows us to identify both grid translation and $ZX$-dualities for the relevant codes. Furthermore,  by allowing unrestricted physical \textit{SWAP} operations we find additional generators of the automorphism group. These typically implement a logical \textit{CNOT} circuit between the primed and unprimed logical blocks (ie without any logical $H$ gate). The logical action of one of these is shown in  \Cref{fig:n108k8d10} for the $[[108,8,10]]$ code. 
In  \Cref{tab:bb_codes} we list the orders of the automorphism and logical action groups we find, and compare these to the results of  \cite{bravyi_high-threshold_2024} and \cite{cross_linear-size_2024}.

\begin{table}[H]
    \centering
    \begin{tabular}{|c|c|c|c|c|}
        \hline 
        Code & $|\text{Aut}(\mathcal{G})|$ &$|\text{Aut}_\text{IBM}|$ & $|\Lambda|$ & $|\Lambda_\text{IBM}|$ \\ 
        \hline
        $[[72,12,6]]$ & 864 & 72 & 864 & 72 \\
        $[[90,8,10]]$ & 360 & 90& 72 & - \\ 
        $[[108,8,10]]$ & 216 & 108& 36  & -\\
        $[[144,12,12]]$ & 288 & 144&144 & 72  \\
        $[[288,12,18]]$ & 1728 &288& 432 & - \\
        $[[360,12,\leq24]]$ & 720& 360 & 144  &  -\\
        \hline
    \end{tabular}
    \caption{Orders of the automorphism and logical action groups of the bivariate bicycle codes of \cite{bravyi_high-threshold_2024}. The order of the automorphism groups we find is denoted $|\text{Aut}(\mathcal{G})|$. The order of the group generated by the grid translation automorphisms and the  ZX-duality identified in \cite{bravyi_high-threshold_2024} is denoted $\text{Aut}_\text{IBM}$ and is the same as the number of physical qubits of the code.  The size of the logical action group we find is denoted $|\Lambda|$. Where available, we compare this to the size of the logical action group found in  \cite{bravyi_high-threshold_2024}, which is denoted  $|\Lambda_{\text{IBM}}|$.}
    \label{tab:bb_codes}
\end{table}
\noindent \href{https://github.com/hsayginel/autqec/tree/main/examples/bivariate_bicycle_codes}{Our results are available as an interactive notebook.}.

\section{Computational Complexity of Finding Automorphism Groups}\label{sec:aut_complexity}

In this section, we discuss the computational complexity of finding automorphism gates using our methods. 
The most computationally complex part is finding the automorphism groups of the related binary linear code.
Finding code automorphisms scales exponentially in the code dimension so is not practical in all cases. 
For larger codes, we instead calculate the matrix automorphisms of the 2- or 3-block binary representations of the stabilizer generators.
Matrix automorphisms can be found by mapping to a graph isomorphism problem and using open source packages which have quasi-polynomial run-time.

\subsection{Code Automorphisms}\label{sec:code_aut_complexity}
In \Cref{sec:auts_of_stab_codes} we considered \textbf{code automorphisms} of a binary linear code with generator matrix $G$.
These are permutations of the bits of the code which result in a permutation of the codewords.
The codewords $\braket{G}$ are the span of $G$ over $\mathbb{F}_2$, and so we write $\text{Aut}(\braket{G})$ for the group of automorphisms of the code with generator matrix $G$.  

Finding code automorphisms is of exponential complexity in the worst case. 
The complexity of Leon's algorithm is exponential in the code dimension \cite{leon_computing_1982}.
Sendrier and Skersys's algorithm has improved efficiency, but is still exponential in the dimension of the code's hull (i.e. the intersection of the code and its dual \cite{sendrierskersys}).

To calculate code automorphisms, we use the proprietary MAGMA computational algebra package \cite{bosma_magma_1997} which has an implementation of Leon's algorithm.
For large codes, the computation time of MAGMA's implementation becomes impractical - for instance we were only able to use MAGMA for bivariate bicycle codes on up to $n=144$ qubits using the $3$ block binary linear code described in \Cref{sec:swap_transversal} and up to $n=360$ using the $2$ block ZX-Duality representation of \Cref{sec:zx_duality} as can be seen in  \Cref{fig:bb_codes_time}. 

\subsection{Matrix Automorphisms}\label{sec:matrix_aut_complexity}
For larger codes, we instead consider \textbf{matrix automorphisms}.
The matrix automorphisms $\text{Aut}(M)$ of the matrix $M$  are permutations of the columns of the matrix which result in a permutation of the rows of $M$.
If $M$ has the same span as the generator matrix ${G}$, then $\text{Aut}(M)$ is a subgroup of $\text{Aut}(\braket{G})$.
On the other hand, if $M$ is the matrix whose rows are the codewords $\braket{G}$, we have $\text{Aut}(M)=\text{Aut}(\braket{G})$.
Matrix automorphisms can be calculated by mapping to a graph isomorphism problem. 
The graph isomorphism problem is thought to be NP-intermediate and quasi-polynomial complexity algorithms are available in the open-source  Bliss and nauty packages \cite{Bliss,nauty}.


To map to a graph isomorphism problem, we start with an $a \times b$ binary matrix $M$ which has the same span as $G$.
The rows of $M$ need not be independent, and may be an over-complete basis.
We consider the automorphisms of the a bi-colored graph $\mathcal{G}_M$ with $a+b$ vertices defined as follows. 
Each row of $M$ corresponds to a white vertex and each column to a black vertex. 
There is an edge between white vertex $i$ and black vertex $j$ if and only if $M_{ij} = 1$.
\textbf{Graph automorphisms} are permutations of the white vertices and/or black vertices which result in a permutation of the edges of the graph.
A generating set of graph automorphisms can be calculated using the open-source nauty or Bliss packages even for very large graphs.

\subsection{Comparison of Matrix and Code Automorphism Methods}\label{sec:code_v_matrix_aut_complexity}
Where $M$ has the same span as $G$, the matrix automorphism group of $M$ is a subgroup of the  automorphism group of the code with generator matrix $G$ (i.e. $\text{Aut}(M) \subseteq \text{Aut}(\braket{G})$).
In general, this is a strict inclusion relationship but in some cases, particularly where we use an over-complete set of stabilizer generators, the groups are equal.

As our first example, we consider the  automorphisms of the $[[5,1,3]]$ code. 
We first consider the independent set of 4 stabilizer generators as set out in Example \ref{eg:513_standard_form}.
Using a 3-block representation (see \Cref{sec:3bit}), we find a trivial  matrix automorphism group. 
If we instead use an over-complete set of stabilizer generators (i.e. the $5$ cyclic shifts of the check $XZZXI$), we find a matrix automorphism group of order $20$.
This corresponds to cyclic shifts of the qubits which act trivially and the \textit{ZX}-duality of Example \ref{eg:513_H+SWAP}.
Finally, if we consider all $2^4 = 16$ elements of the  stabilizer group, we obtain a matrix automorphism group of order $360$ which matches the order of the code automorphism group.

\begin{figure}[H]
    \centering
    \includegraphics[width=\linewidth]{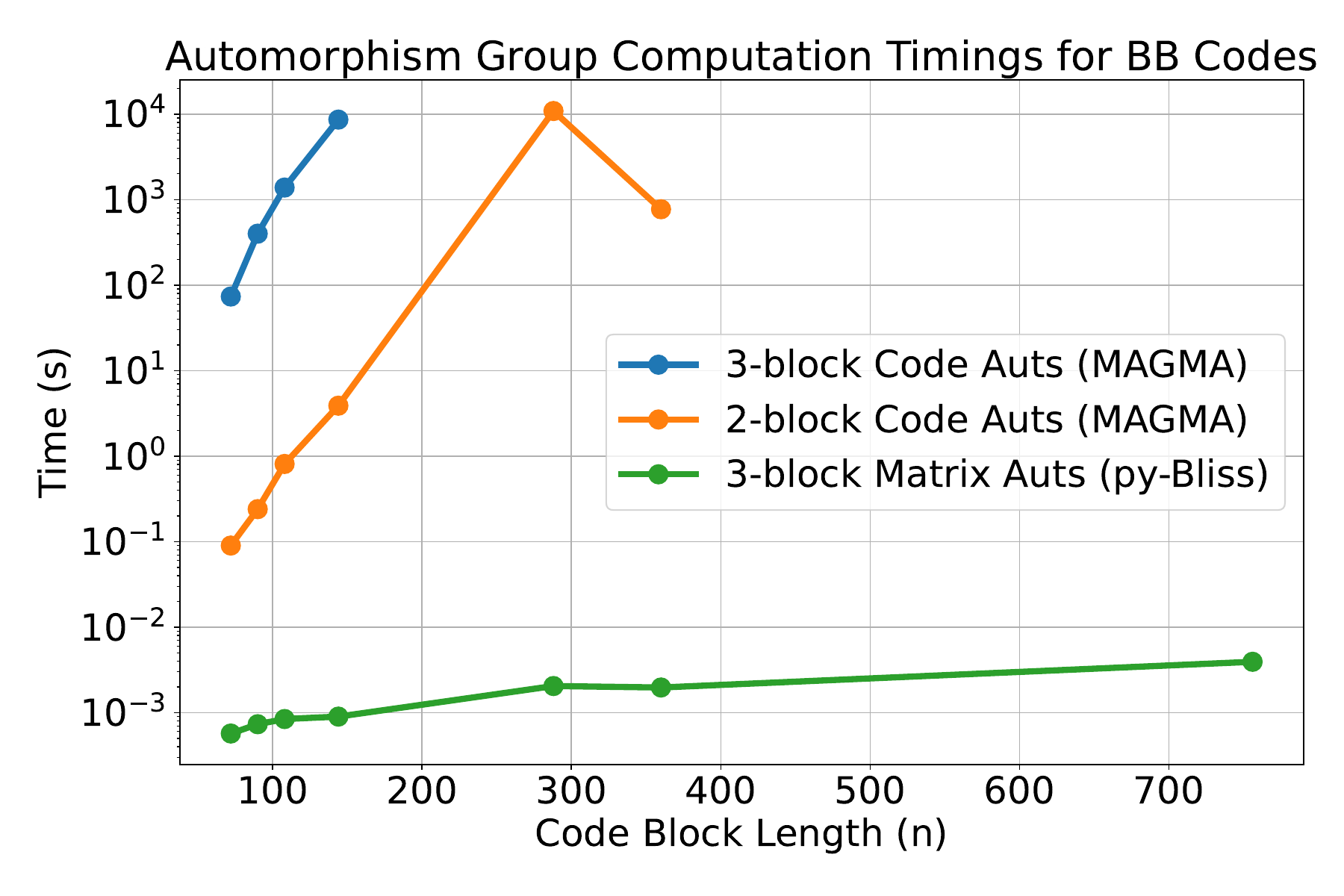}
    \caption{Time analysis for computing automorphisms of bivariate bicycle codes. Blue graph: time taken to compute code automorphisms using MAGMA for the 3-block representation in  \Cref{tab:aut_family} Row 4. Orange graph: time taken to compute code automorphisms using MAGMA for the 2-block representation in  \Cref{tab:aut_family} Row 1. Green line: time taken to compute matrix automorphisms for the 3-block representation in  \Cref{tab:aut_family} Row 4. Note that the group orders found via each representation are the same for all BB codes.}
    \label{fig:bb_codes_time}
\end{figure}

In the case of bivariate bicycle codes, matrix automorphisms may suffice to generate the entire code automorphism group. 
bivariate bicycle codes have a canonical over-complete set of $n$ stabilizer generators.
The order of the matrix automorphism group of the canonical stabilizers matches the order of the code automorphism group for members of the Bivariate Bicycle Code family of \cite{bravyi_high-threshold_2024} for up to $n=360$ qubits.
We conjecture that the canonical stabilizer generators form a complete set of minimum-weight stabilizers for bivariate bicycle codes, though proving this is an open question.
As automorphisms are weight-preserving, the set of minimum-weight stabilizer generators are closed under the action of code automorphisms, so any code automorphism is also an automorphism of the minimum-weight stabilizer generators.
As illustrated in \Cref{fig:bb_codes_time}, matrix automorphisms of bivariate bicycle codes take significantly less time to find than code automorphisms.

\section{Conclusion}\label{sec:conclusion}
In this work, we presented a number of different algorithms to identify the logical Clifford gates implementable through code automorphisms. The key to our methods is mapping a stabilizer code to a binary linear code. We first extend the methods for identifying fold-transversal \textit{SWAP} + $H$ operations to non-CSS stabilizer codes. We also generalize our method to finding \textit{SWAP} + $S$ and \textit{SWAP} + $\sqrt{X}$ physical circuits. Furthermore, we combined the above to find all single-qubit Clifford + \textit{SWAP} circuits preserving the codespace, using a $3$-block representation of the stabilizer code. These correspond to  code automorphisms over $GF(4)$ as set out in \cite{calderbank1997}, but we more rigorously apply Pauli corrections to ensure the signs of the stabilizer generators are preserved and interpret the logical actions of the resulting gates.
Our methods show some very interesting results when applied to the best-distance codes with $1$ logical qubit. Specifically, we find that most of them allow for \textit{SWAP}-transversal implementations of the single-qubit Clifford group, far more than what is achievable using only transversal gates. Finally, we found that for the bivariate bicycle codes \cite{bravyi_high-threshold_2024}, there exist further automorphisms with interesting logical actions if we allow for arbitrary physical qubit \textit{SWAP}s. We provide a package which implements all the aforementioned methods using Python and either the computational algebra software MAGMA or the open-source Bliss package for computing automorphism groups. 

Overall, our results show that \textit{SWAP} gates can substantially enrich the logical gate sets implementable in stabilizer codes.

\subsection{Open Questions and Future Work}
Several follow-up questions naturally arise. First, we observe that automorphisms play an increasingly important role in modern fault-tolerant protocols—particularly in lattice surgery and extractor systems \cite{cross_linear-size_2024, he_extractors_2025} where they can reduce the auxiliary qubit overhead. Our automated framework discovers symmetries beyond those previously identified analytically, raising the question of whether these additional automorphisms can lead to further resource savings in such protocols. Second, connections to decoding are also compelling: recent work \cite{Koutsioumpas2025AutomorphismED} has shown that codes with rich automorphism groups admit highly parallelizable, accurate, and efficient decoding algorithms. These observations underscore the practical value of symmetries in quantum error correction codes and motivate further exploration into their role in both decoding and logical gate construction. Finally, we note that our proposed constructions may be especially well-suited for experimental platforms with native \textit{SWAP} gates, and we hope this work can contribute to accelerating near-term demonstrations of fault-tolerant gates.

\begin{acknowledgments}
HS is supported by the Engineering and Physical Sciences Research Council [grant number EP/S021582/1]. HS also acknowledges support from the National Physical Laboratory. MW and DEB are supported by the Engineering and Physical Sciences Research CouncilEP/W032635/1]. AR, MW and DEB are supported by the Engineering and Physical Sciences Research Council [grant number EP/S005021/1]. SK and DEB are supported by the Engineering and Physical Sciences Research Council [grant number EP/Y004620/1]. SK, MW, and DEB are supported by the Engineering and Physical Sciences Research Council [grant number EP/T001062/1].
The authors would like to thank Geoff Bailey, Simon Burton, Earl Campbell, Alastair Kay and Heather Leitch for helpful discussions. 
\end{acknowledgments}

\bibliography{bibliography}

\clearpage
\onecolumngrid
\appendix

\section{Correspondence of Permutations and Symplectic Matrices}\label{sec:auts2stabilizercode}
We will now give a more detailed description of our approach to mapping automorphisms of a binary linear code to \textit{SWAP}-transversal Clifford gates. We begin by showing a mapping of the circuits to automorphisms of a binary linear code. We then construct the mapping of these permutations into symplectic matrices using the transformation matrix.

\subsection{Connection between \textit{SWAP}-Transversal Operations and Automorphism Groups of Binary Cyclic Block Codes}\label{sec:aut_grp_intersection}
We will now prove that there is a 1-to-1 correspondence between the automorphism group of the classical binary linear code of size $3n$ generated by the matrix:

\[
B_n:= \left[
\begin{array}{c|c|c}
I_n & I_n & I_n\\
\end{array}
\right].
\] and all circuits composed of $H,S$ and \textit{SWAP} gates on $n$ qubits, modulo Pauli operators.

We start by finding a description of the automorphism group of the binary linear code $\braket{B_n}$ generated by $B_n$. Let $e_i$ be the binary vector of length $n$ with $1$ in position $i$ and $0$ everywhere else.
The generator matrix of $\braket{B_n}$ can be written as:
\begin{equation}
    g_i:=(e_i|e_i|e_i),  \forall 1\leq i\leq n.
\end{equation}

Any codeword $b\in\braket{B_n}$ will thus be of the form: 
\begin{equation}
    b=\sum_{i=1}^{n}a_i g_i=(a_1,a_2,...,a_n,a_1,a_2,...,a_n,a_1,a_2,...,a_n)
\label{eq:3bcodewordform}\end{equation}
for some $a_i \in \{0,1\}$. 
\begin{theorem}\label{thm:3blockblc}
The automorphism group of the binary linear code generated by $B_n$,  $Aut(\braket{B_n})$ is generated by permutations of the following three types:
\begin{enumerate}
    \item $(i,n+i), \forall 1\leq i \leq n$
    \item $(n+i,2n+i), \forall 1\leq i \leq n$
    \item $(i,j)(n+i,n+j)(2n+i,2n+j), \forall 1\leq i< j \leq n $
\end{enumerate} and its size is:
    $|Aut(\braket{B_n})|=6^n n!$
\end{theorem}
\textbf{Proof}
    
Let $P$ be an automorphism of $\braket{B_n}$ ie $P\in Aut(\braket{B_n})$. As we described in \Cref{sec:auts_of_stab_codes}, $P$ will be a permutation of the $3n$ bits,  $P \in S_{3n}$, that preserves the codespace.  We check the action of $P$ on a generator $g_i$. As this is a binary vector of length $3n$ with $3$ non-zero coordinates, $P(g_i)$ will also be a binary vector of length $3n$ with $3$ non-zero coordinates. There are two cases depending on where the mapped non-zero coordinates will be: 
we can either have more than $2$ non zero elements in one of the three $n-$ tuples, or exactly $1$ non zero element in each $n$ tuple.

Specifically, we have the following cases:

\begin{enumerate}
    \item $P(g_i) = (e_j+e_k|e_l|0) $ or $P(g_i) = (e_j+e_k+e_l|0|0) $ for some $1\leq j,k,l \leq n.$ This is not a valid codeword of $\braket{B_n}$, as it is not of the form in  \Cref{eq:3bcodewordform}. Similarly if we have at least $2$ non zero elements in the second or third tuple. 

    \item  $P(g_i) = (e_j|e_k|e_l)$ for some $1\leq j,k,l \leq n.$ If $j \neq k$, then matching the form of \Cref{eq:3bcodewordform}, from the first $n$ bits we get that $a_j=1, a_k=0$ while from the second $n$ bits we find that $a_j=0,a_k=1$ which is a contradiction. Similarly, if $j \neq l$.   Hence, we require $j=k=l$, and  $P(g_i) = (e_j|e_j|e_j)=g_j$. This is a valid codeword of $\braket{B_n}$ and hence $P$ is a valid automorphism.

\end{enumerate}

We see that all the automorphisms of $\braket{B_n}$, are of the form $P(g_i) = g_j$ for some $1\leq i,j \leq n$, or equivalently $P: (e_i|e_i|e_i)\mapsto (e_j|e_j|e_j).$ Thus, for each $i$, $P$ can swap column $i$ with $n+i$ or $2n+i$, or column $n+i$ with $2n+i$, and then columns $i$ with $j$, $n+i$ with $n+j$ and $2n+i$ with $2n+j$. This is equivalent to permuting each column to the same index in each of the $3$ blocks and permuting $n$ indices separately, thus the elements of $Aut(\braket{B_n})$ are generated by the following permutations:
\begin{itemize}
    \item $(i,n+i), \forall 1\leq i \leq n$
    \item $(n+i,2n+i), \forall 1\leq i \leq n$
    \item $(i,j)(n+i,n+j)(2n+i,2n+j), \forall 1\leq i< j \leq n $
\end{itemize}

Hence the order of the automorphism group of the binary linear code $\braket{B_n}$  is: 
\begin{equation}
    |Aut(\braket{B_n})|=|S_3|^n|S_n|=6^n n!.
\end{equation}\null\hfill$\Box$

Let $K$ be the set of all possible circuits composed of $H,S$ and \textit{SWAP}s. These form a group with group operation being circuit composition.

We now introduce the map $\phi: Aut(\braket{B_n}) \mapsto K$ defined by:
\begin{align}
    &(i,n+i)\mapsto H_i, &\forall 1\leq i \leq n\\
    &(n+i,2n+i)\mapsto S_i, &\forall 1\leq i \leq n\label{eq:Scycle}\\
    &(i,j)(n+i,n+j)(2n+i,2n+j)\mapsto \text{\textit{SWAP}}_{ij}, &\forall 1\leq i\leq j \leq n. 
\end{align}

This is a group isomorphism, as multiplying the permutations before applying the map has the same effect as applying $\phi$ and then composing the circuits.

Similarly, for the $2$ block representations discussed in  \Cref{sec:zx_duality}, the automorphism group of \[
B:= \left[
\begin{array}{c|c}
I_n & I_n \\
\end{array}
\right].
\] is generated by permutations of the type $(i,n+i)$ and $(i,j)(n+i,n+j)$ corresponding to a transversal Clifford gate, such as $H_i$ for the symplectic binary representation, and $\text{\textit{SWAP}}_{ij}$ respectively. By the same argument as in  \Cref{thm:3blockblc}, the size of the group is then $|S_2|^n|S_n|=2^n n!$.

\subsection{Mapping of Permutation Automorphisms to Symplectic Matrices}\label{sec:perm2sym}
We now show how each of the allowed permutation automorphisms translate to symplectic matrices under conjugation by the transformation matrix $\mathcal{E}$.
For the 3-block representation $[G_X|G_Z|G_X\oplus G_Z]$  of  \Cref{sec:swap_transversal}, the transformation matrix and its inverse are given by:
\begin{align}
\mathcal{E} := \begin{bmatrix}I&0&I\\0&I&I\\I&I&I\end{bmatrix}; \mathcal{E}^{-1} := \begin{bmatrix}0&I&I\\I&0&I\\I&I&I\end{bmatrix}.
\end{align}

\subsubsection{S-Type Permutation}
On a single physical qubit, the permutation corresponding to an $S$ gate as we defined it above in \Cref{eq:Scycle} is $(2,3)$ in cycle notation. Its matrix representation, thus is given by:
\begin{align}
P_S := \begin{bmatrix}1&0&0\\0&0&1\\0&1&0\end{bmatrix}.
\end{align}
We see that right multiplication by this matrix swaps the Z and XZ component of the $3$ block binary representation of \Cref{sec:3bit}.

We calculate the corresponding symplectic matrix as follows:
\begin{align}U_S \oplus W_S &:= \mathcal{E}P_S\mathcal{E}^{-1}\\
&= \begin{bmatrix}1&0&1\\0&1&1\\1&1&1\end{bmatrix}
\begin{bmatrix}1&0&0\\0&0&1\\0&1&0\end{bmatrix}
\begin{bmatrix}0&1&1\\1&0&1\\1&1&1\end{bmatrix}\\
&= 
\begin{bmatrix}1&1&0\\0&1&1\\1&1&1\end{bmatrix}
\begin{bmatrix}0&1&1\\1&0&1\\1&1&1\end{bmatrix}\\
&= 
\begin{bmatrix}1&1&0\\0&1&0\\0&0&1\end{bmatrix}\\
&=
\begin{bmatrix}1&1\\0&1\end{bmatrix}\oplus \begin{bmatrix}1\end{bmatrix}.\end{align}
The symplectic matrix $U_S$ corresponds to an $S$ operator (see  Example \ref{eg:symplectic}), and the action of $W_S$ on the $XZ$ block is trivial.

\subsubsection{Hadamard-Type Permutation}
The permutation matrix swapping the X and Z blocks on a single-qubit is given by:
\begin{align}
P_H := \begin{bmatrix}0&1&0\\1&0&0\\0&0&1\end{bmatrix}
\end{align}
Under conjugation by $\mathcal{E}$, we obtain the following symplectic matrix:
\begin{align}U_H \oplus W_H &:= \mathcal{E}P\mathcal{E}^{-1}
=\begin{bmatrix}0&1\\1&0\end{bmatrix}\oplus \begin{bmatrix}1\end{bmatrix}\end{align}
The symplectic matrix $U_H$ corresponds to a single-qubit $H$ operator (see Example \ref{eg:symplectic}).

\subsubsection{SWAP-Type Permutation}
Now consider a permutation of qubits given by an $n \times n$ permutation matrix $Q$. The corresponding permutation matrix in the 3-block representation is given by
\begin{align}
P_Q:= \begin{bmatrix}Q&0&0\\0&Q&0\\0&0&Q\end{bmatrix}
\end{align}
Under conjugation by $\mathcal{E}$, we obtain the following symplectic matrix: 
\begin{align}U_Q \oplus W_Q &:= \mathcal{E}P\mathcal{E}^{-1} = P_Q = \begin{bmatrix}Q&0\\0&Q\end{bmatrix}\oplus \begin{bmatrix}Q\end{bmatrix}\end{align}
The symplectic matrix $U_Q$ corresponds to the permutation $Q$ on the qubits. In this case, $W_Q$ is not trivial and implements a corresponding permutation on the $XZ$ block.

\section{Decomposition of Symplectic Matrix into Circuit Form}\label{sec:clifford_decomp_proof}
In this Appendix, we show how to decompose a symplectic matrix into a circuit involving $H, CNOT, C(X,X), \sqrt{X}, CZ$ and $S$ gates.
We first set out our main theorem which shows that such a decomposition exists, then give an algorithm for obtaining the decomposition in practice.

\subsection{Existence of Circuit Decomposition}\label{sec:circuit_decomp_existence}

In this section, we show that any symplectic matrix can be written as a product of four symplectic matrices - one corresponding to $C(X,X)$ and $\sqrt{X}$ gates, one corresponding to \textit{CZ} and $S$ gates, one corresponding to \textit{CNOT} gates and one corresponding to $H$ gates.

\begin{theorem}[Product form of symplectic matrix]\label{thm:circuit}
Any symplectic matrix can be written in the form $U = U_AU_BU_CU_H$ where:
\begin{itemize}
    \item $U_A:=\left[\begin{array}{c|c}I&0\\\hline A&I\end{array}\right]$ for some symmetric $n\times n$ binary matrix $A$ is a set of $C(X,X)$ and $\sqrt{X}$ gates;
    \item $U_B:=\left[\begin{array}{c|c}I&B\\\hline 0&I\end{array}\right]$ for some symmetric $n\times n$ binary matrix $B$ is a set of CZ and S gates; 
    \item $U_C := \left[\begin{array}{c|c}C&0\\\hline 0&C^{-T}\end{array}\right]$ for some invertible $n\times n$ binary matrix $C$ is a set of CNOT gates; and
    \item $U_H := \left[\begin{array}{c|c}\text{diag}(1-\mathbf{h})&\text{diag}(\mathbf{h})\\\hline\text{diag}(\mathbf{h})&\text{diag}(1-\mathbf{h})\end{array}\right]$ for some length $n$ binary vector $\mathbf{h}$  corresponding to a set of $H$ gates.
\end{itemize}
\end{theorem}
\textbf{Proof}
To solve for $U_H$, we follow the method outlined in Lemma 6 of \cite{aaronson_improved_2004}. Let $U := \left[\begin{array}{c|c}C_0&B_0\\\hline A_0&C_1\end{array}\right]$. The rows of $[C_0|B_0]$ are independent so if $C_0$ is of rank $r < n$, then for some $B_1, C_2$ and $B_2$ of rank $n-r$ we have:
\begin{align}
    RREF(C_0|B_0) = \left[\begin{array}{c|c}C_2&B_1\\0&B_2\end{array}\right].
\end{align}
Now calculate $RREF(B_2)$ and form the binary vector $\mathbf{h}$ of length $n$ where the $i$th component of $\mathbf{h}$ is 1 if and only if $i$ is a pivot when calculating $RREF(B_2)$. Following the reasoning of Lemma 6 of \cite{aaronson_improved_2004}, the upper left hand block of $UU_H$ is invertible so we set this to be $C$ and form $U_C$. Applying the inverse  $U_C^{-1} := \left[\begin{array}{c|c}C^{-1}&0\\\hline 0&C^{T}\end{array}\right]$ to the RHS of $UU_H$, for some $A, B, C_3$ we have:
\begin{align}
    UU_HU_C^{-1} = \left[\begin{array}{c|c}I&B\\\hline A&C_3\end{array}\right].
\end{align}
This matrix is symplectic so 
\begin{align}
    \Omega = \left[\begin{array}{c|c}0&I\\\hline I&0\end{array}\right] &= \left[\begin{array}{c|c}I&B\\\hline A&C_3\end{array}\right] \Omega \left[\begin{array}{c|c}I&B\\\hline A&C_3\end{array}\right]^T\\ &=\left[\begin{array}{c|c}B&I\\\hline C_3&A\end{array}\right] \left[\begin{array}{c|c}I&A^T\\\hline B_T&C_3^T\end{array}\right]\\&=\left[\begin{array}{c|c}B+B^T&BA^T+C_3^T\\\hline C_3+AB^T&C_3A^T+AC_3^T\end{array}\right].
\end{align}
Hence $B + B^T = 0$ and so $B$ is symmetric. The transpose of $UU_HU_C^{-1}$ is also symplectic and this implies that $A$ is symmetric as well. We construct $U_A$ and $U_B$, noting that these are self-inverse, and apply them to the RHS:
\begin{align}
    UU_HU_C^{-1}U_BU_A &= \left[\begin{array}{c|c}I&B\\\hline A&C_3\end{array}\right]\left[\begin{array}{c|c}I&B\\\hline 0&I\end{array}\right]\left[\begin{array}{c|c}I&0\\\hline A&I\end{array}\right]\\&=\left[\begin{array}{c|c}I&B\\\hline A&C_3\end{array}\right]\left[\begin{array}{c|c}I+BA&B\\\hline A&I\end{array}\right]\\&=\left[\begin{array}{c|c}I+BA+BA&B+B\\\hline A+ABA+C_3A&AB+C_3\end{array}\right]\\&=\left[\begin{array}{c|c}I&0\\\hline (I+AB+C_3)A&AB+C_3\end{array}\right].
\end{align}
This is a symplectic matrix and this implies that $AB+C_3 = I$ because, writing $D:=(I+AB+C_3)A$ and $E:= AB+C_3$:
\begin{align}
    \Omega = \left[\begin{array}{c|c}0&I\\\hline I&0\end{array}\right] &=\left[\begin{array}{c|c}I&0\\\hline D&E\end{array}\right] \Omega\left[\begin{array}{c|c}I&0\\\hline D&E\end{array}\right] ^T\\&=\left[\begin{array}{c|c}0&I\\\hline E&D\end{array}\right] \left[\begin{array}{c|c}I&D^T\\\hline 0&E^T\end{array}\right] \\&=\left[\begin{array}{c|c}0&E^T\\\hline E&ED^T+DE^T\end{array}\right].
\end{align}
This in turn implies that $UU_HU_C^{-1}U_BU_A = I$ and so $U = U_AU_BU_CU_H$ as claimed.\null\hfill$\Box$

\subsection{Calculating the Circuit Decomposition}\label{sec:circuit_decomp_alg}
We now show how to efficiently calculate the circuit decomposition in practice for a symplectic matrix $U$.
We first calculate $\mathbf{h}$ and $U_H$ as described in  \Cref{thm:circuit} such that:
\begin{align}
    UU_H &= U_AU_BU_C\\&=\left[\begin{array}{c|c}I&0\\\hline A&I\end{array}\right]\left[\begin{array}{c|c}I&B\\\hline 0&I\end{array}\right]\left[\begin{array}{c|c}C&0\\\hline 0&C^{-T}\end{array}\right]\\&= \left[\begin{array}{c|c}I&B\\\hline A&I+AB\end{array}\right]\left[\begin{array}{c|c}C&0\\\hline 0&C^{-T}\end{array}\right]\\&= \left[\begin{array}{c|c}C&BC^{-T}\\\hline AC&(I+AB)C^{-T}\end{array}\right].
\end{align}
This matrix form gives $C$ directly. We obtain $A$ and $B$ from the matrix form as follows:
\begin{align}
    A&:= (AC)C^{-1};\\
    B&:=(BC^{-T})C^T.
\end{align}

\section{Algorithms}

In this section, we set out key algorithms in more detail. 
We first describe our method for constructing a related binary linear code from a given stabilizer code, finding the automorphisms of the code and then constructing the corresponding symplectic matrices acting on the original codespace. We then describe the algorithm for finding a Pauli correction to ensure that the phases of the stabilizer group are preserved. Finally, we describe an algorithm for checking if a desired logical action can be implemented from a set of generating logical operators.

\begin{algorithm}[H]
\caption{Automorphisms of a binary matrix representation of a stabilizer code.}
\KwIn{
\begin{enumerate}
    \item 
Representation of the stabilizer generators as the binary matrix $G:=\left[\begin{array}{c|c}G_X&G_Z\end{array}\right]$; and
\item Transformation matrix $\mathcal{E}$ corresponding to either a 2 or 3 block binary representation.
\end{enumerate}
}
\KwOut{A set of symplectic matrices $\bar{U}_\text{S}$ which generate the relevant group of Clifford automorphisms.}
\SetKwProg{Method}{Method}{:}{}
\Method{}{
    \begin{enumerate}
        \item Form the binary matrix $G_\mathcal{E} :=G\mathcal{E}$ for 2-block representations or $G_\mathcal{E} := (G |0) \mathcal{E}$ for 3-block representations;
        \item Let $B:=\begin{bmatrix}I&I\end{bmatrix}$ for 2-block representations or $B:=\begin{bmatrix}I&I&I\end{bmatrix}$ for 3-block representations;
        \item Find the generators of $\text{Aut}(\braket{G_\mathcal{E}}) \cap \text{Aut}(\braket{B})$  where $\braket{B}$ is the binary linear code with generator matrix $B$, using a computational algebra program e.g. MAGMA;
        \item Each generator of $\text{Aut}(\braket{G_\mathcal{E}}) \cap \text{Aut}(\braket{B})$ is a column permutation and has a matrix representation $P$. For two-block representations, the symplectic matrix  $\bar{U}_\text{SP} := \mathcal{E} P \mathcal{E}^{-1}$; for 3-block representations we have a direct sum  $\mathcal{E} P \mathcal{E}^{-1} = \bar{U}_\text{S} \oplus W$;
        \item Return the symplectic matrices $\bar{U}_\text{S}$ for each generator $P$ of $\text{Aut}(\braket{G_\mathcal{E}}) \cap \text{Aut}(\braket{B})$.
    \end{enumerate}}
\end{algorithm}

\begin{algorithm}[H]
\caption{Pauli Correction and Logical Action of a Clifford operator.}
\KwIn{
\begin{enumerate}
    \item A tableau $\tau := \begin{bmatrix}
        G\\L_X\\R\\L_Z
    \end{bmatrix}$ for an $[[n,k]]$ stabilizer code such that $\tau \Omega \tau = \Omega$ (see  \Cref{sec:tableau});
    \item A Clifford circuit $\bar{U}_S$ acting on $n$ qubits;
\end{enumerate}}
\KwOut{
\begin{enumerate}
    \item If $\bar{U}_S$ is not a logical operator, return FALSE. \item Otherwise, return a product of Pauli operators $\bar{U}_P$ which ensures that the signs of the stabilizer generators are preserved by $\bar{U}:=\bar{U}_S\bar{U}_P$, plus a $2k\times 2k$ symplectic matrix $U_\text{ACT}$  which describes the logical action of $\bar{U}$.
    \end{enumerate}}
\SetKwProg{Method}{Method}{:}{}
\Method{}{
    \begin{enumerate}
        \item Set $\bar{U}_P:=I$ and set $U_\text{ACT}$ to be an empty list; 
        \item For each row $(\mathbf{x}_i|\mathbf{z}_i)$ of $\tau$ corresponding to a stabilizer, logical Pauli generator  $i^{p_i}X(\mathbf{x}_i)Z(\mathbf{z}_i)$, calculate the mapped operator $\bar{U}_S \left(i^{p_i} X(\mathbf{x}_i)Z(\mathbf{z}_i)\right) \bar{U}_S^\dag := i^u X(\mathbf{x}')Z(\mathbf{z}')$ using the update rules of  \Cref{tab:clifford_actions};
        \item Let $\mathbf{b}:=(\mathbf{x}' | \mathbf{z}')\Omega \tau^T \Omega=(\mathbf{g} | \mathbf{a}_X | \mathbf{r}|\mathbf{a}_X)$ as in  \Cref{eq:b_vector}. This is a binary string of length $2n$ such that that $(\mathbf{x}' | \mathbf{z}') = \mathbf{b}\tau = \mathbf{g}G + \mathbf{a}_X L_X +\mathbf{r}R+\mathbf{a}_Z L_Z$;
        \item If $\mathbf{r}$ is non-zero, then the mapped operator is the product of at least one destabilizer and so $\bar{U}_S$ does not preserve the codespace. Return FALSE; 
        \item If row $i$ corresponds to a stabilizer and $(\mathbf{a}_X|\mathbf{a}_Z)$ is non-zero, then the mapped operator is the product of at least one logical Pauli generator and $\bar{U}_S$ does not preserve stabilizers. Return FALSE; 
        \item If row $i$ corresponds to a logical Pauli, append the row $(\mathbf{a}_X|\mathbf{a}_Z)$ to $U_\text{ACT}$. This is a vector of length $2k$ corresponding to the logical Pauli generators appearing in the mapped logical Pauli; 
        \item Calculate the product $i^{-\mathbf{a}_X\cdot \mathbf{a}_Z}\prod_{0 \le j < 2n}\big(i^{p_j}X(\mathbf{x}_j)Z(\mathbf{z}_j)\big)^{\mathbf{b}[j]} := i^v X(\mathbf{x}')Z(\mathbf{z}')$, noting the phase adjustment $i^{-\mathbf{a}_X\cdot \mathbf{a}_Z}$;
        \item Let $h := (j + n ) \mod 2n$. If $u \ne v \mod 4$, multiply $\bar{U}_P$ by the anti-commuting operator $X(\mathbf{x}_h)Z(\mathbf{z}_h)$;
        \item Return $U_\text{ACT},\bar{U}_P$.
    \end{enumerate}}\label{alg:pauli_correction}
\end{algorithm}

\begin{algorithm}[H]
\caption{Determining whether a logical operator can be implemented through automorphisms of the code.}\label{alg:lo_search}
\KwIn{}
\begin{enumerate}
    \item Desired Clifford Logical operator ${U}$ acting on $k$ logical qubits of the code.
    \item Logical Clifford operators $U_\text{ACT}$ implemented by the automorphism group generators of the chosen stabilizer code representation.
\end{enumerate}
\KwOut{If ${U}$ is in the group of logical operators generated by the action of the physical circuits obtained through automorphisms of the chosen representation, return a physical circuit 
$\bar{U}$ implementing it. Otherwise return FALSE.}
\SetKwProg{Method}{Method}{:}{}
\Method{}{
    \begin{enumerate}
        \item Write ${U}$ as an element of $\text{Sp}_{2k}$, the group of $2k \times 2k$ symplectic matrices.
        \item For each generator $a_i, 1\leq i \leq l$, of the automorphism group of the chosen stabilizer representation, obtain its logical action $U_{\text{ACT}_i}$ with the correct phases from  \Cref{alg:pauli_correction}. This is a $2k \times 2k$ symplectic matrix.
        \item Form the group $\Lambda:=\langle U_{\text{ACT}_1},\dots , U_{\text{ACT}_l}\rangle \subseteq \text{Sp}_{2k}$. 
        \item Check whether $U\in \Lambda$. If not, return FALSE and terminate. 
        \item If $U\in \Lambda$, find its decomposition in terms of generators $U_{\text{ACT}_i}$ through a presentation of $\Lambda$. 
        \item Run  \Cref{alg:pauli_correction} again to find any potential Pauli corrections $\bar{U}_P$ to the above decomposition of $U$.
        \item Return the composition of the physical circuits $\bar{U_i}$ corresponding to each $U_{\text{ACT}_i}$ in the decomposition of $U$ and any Pauli corrections $\bar{U}_P$. This circuit implements a logical $U$. 
    \end{enumerate}}
\end{algorithm}

\section{Embedded Codes}\label{app:embedded_codes}
The methods described in  \Cref{sec:auts_of_stab_codes} are able to identify logical operators which use \textit{SWAP} and single-qubit physical gates.
The embedded code technique allows us to identify a wider range of logical operators which  also use two-qubit \textit{CNOT} and \textit{CZ} physical gates.
In this section, we set out the key results for constructing embedded codes and finding logical Clifford operators via permutation automorphisms of a related binary linear code.
The structure of this appendix is as follows. We first describe the embedding operator which is a Clifford unitary used to embed a stabilizer code into a larger Hilbert space. 
We then describe how to calculate the stabilizer generators of the embedded code by applying the embedding operator to Pauli operators. 
Finally, we show how to interpret the automorphisms of the embedded code as logical operators in the original codespace.

\subsection{Embedding Operator Definition}\label{sec:embedding_operator}
Let $M$ be a $m \times n$ binary matrix and let $V:=\begin{bmatrix}I\\M\end{bmatrix}$.
The \textbf{embedding operator} $\mathcal{E}_V$ on $n+m$ qubits is defined as follows.
Let $\mathbf{v}_j$ be row of $V$ and $j \ge n$. Let $i \prec \mathbf{v}_j$ denote that $i$ is in the support of $\mathbf{v}_j$. Define 
\begin{align}
    U_j &:= \prod_{i \prec \mathbf{v}_j} \textit{CNOT}_{ij}; \text{ and}\\
    \mathcal{E}_V &:= \prod_{n \le j < n+m} U_j.
\end{align}
Note that $\mathcal{E}_V$ is a Clifford unitary with $\mathcal{E}_V^{-1} = \mathcal{E}_V$. 
All terms $\textit{CNOT}_{ij}$ in the $U_j$ commute because control qubits are never updated by the relevant operators, and as a result the $U_j$ also commute.
We can for instance select $V$ to be of form $V :=\begin{bmatrix}I_n\\M^n_2\end{bmatrix}$ where $M^n_2$ is the matrix whose $m:=\binom{n}{2}$ rows are the binary strings of length $n$ and weight $2$. In the example below, we see that this corresponds to adding auxiliary qubits which represent pairs of our original qubits.
\begin{example}\label{eg:embedded_code_CNOT}
    Let $n:=2$, $V := \begin{bmatrix}1&0\\0&1\\1&1\end{bmatrix}$ so that $\mathbf{v}_2 = 11$. In this case, $\mathcal{E}_V = U_2 = \textit{CNOT}_{02}\textit{CNOT}_{12}$. Here we calculate the action of $U_2$ on a computational basis vector of form $\ket{x_0,x_1,0}$. We have:
\begin{align*}
    U_2\ket{x_0,x_1,0} &= \textit{CNOT}_{02}\textit{CNOT}_{12}\ket{x_0,x_1,0}\\
    &=\textit{CNOT}_{02}\ket{x_0,x_1,x_1}\\
    &=\ket{x_0,x_1,x_0\oplus x_1}.
\end{align*}
Hence, we can think of qubit 2 as the ‘sum’ of qubits 0 and 1, providing we initialise qubit 2 in the $\ket{0}$ state.
\end{example}
\subsection{Embedded Code Definition}\label{sec:embedded_code_def}
Let $\mathbf{G}$ be a set of independent commuting Pauli operators on $n$ qubits which define a stabilizer code (see \cite{gottesman_stabilizer_1997}).
For a given binary matrix $V := \begin{bmatrix}I\\M\end{bmatrix}$, we  extend the stabilizer code defined by the stabilizer generators $\mathbf{G}$ on $n$ qubits to a stabilizer code on $m+n > n$ qubits by adding stabilizers $Z_j$ for  $n \le j < n+m$ .  Adding the stabilizers $Z_j$ is equivalent to adding auxiliary qubits initialized in the state $\ket{0}$.
The stabilizers of the \textbf{embedded code} are as follows:
\begin{align}
    \mathbf{G}_V := \{\mathcal{E}_V A \mathcal{E}_V: A \in \mathbf{G}\} \cup \{\mathcal{E}_V Z_j \mathcal{E}_V: n \le j <n+m\}.\label{eq:embedded_stabilizers}
\end{align}

\subsection{Action of Embedding Operator}\label{sec:embedding_operator_action}
We now look at the action under conjugation of the embedding operator  $\mathcal{E}_V$ on Pauli operators - this allows us to explicitly calculate the stabilizer generators $\mathbf{G}_V$ of the embedded code. 

For \textbf{Pauli $Z$ operators}, the action of $U_j$ is trivial unless the $Z$ operator acts on qubit $j$. This is because $Z$ operators commute with $\textit{CNOT}$ operators, unless they act on the target of the $\textit{CNOT}$ operator:
\begin{align}
    U_j Z_k U_j =Z_k\left(Z(\mathbf{v}_j)^{\delta_{jk}}\right)=\begin{cases} Z_kZ(\mathbf{v}_j): k = j\\Z_k:\text{ otherwise.}\end{cases}
\end{align}
For \textbf{Pauli $X$ operators}, the action of $U_j$ is trivial unless the $X$ operator acts on a qubit in the support of $\mathbf{v}_j$. This is because $X$ operators commute with $\textit{CNOT}$ operators, unless they act on the control of the $\textit{CNOT}$ operator:
\begin{align}
U_j X_k U_j  = X_k\left(X_j^{\mathbf{v}_j[k]}\right)=  \begin{cases}X_kX_j : k \prec \mathbf{v}_j\\X_k: \text{ otherwise.}\end{cases}\label{eq:embedded_Z}
\end{align}
For a string of Pauli X operators $X(\mathbf{x})$ for $\mathbf{x}$ a binary matrix of length $n$, the action of $\mathcal{E}_V$ is given by $\mathcal{E}_V\big(X(\mathbf{x})\big) = X(\mathbf{x}V^T)$.  To see this, we first note that:
\begin{align*}
    U_jX(\mathbf{x})U_j &= \prod_{0\le k <n}\Big( U_jX_k^{\mathbf{x}[k]}U_j\Big) \\
    &= \prod_{0\le k <n} \Big(X_k^{\mathbf{x}[k]} X_j^{\mathbf{x}[k]\mathbf{v}_j[k]}\Big)\\
    &= X(\mathbf{x})X_j^{\mathbf{x}\cdot \mathbf{v}_j}.
\end{align*}
And so:
\begin{align*}
\mathcal{E}_V\big(X(\mathbf{x})\big) &= \big(\prod_{n \le j < n+m} U_j\big)X(\mathbf{x})\big(\prod_{n \le j < n+m}U_j\Big) \\
&=X(\mathbf{x})\prod_{n \le j < n+m} X_j^{\mathbf{x}\cdot \mathbf{v}_j} \\
&=   X(\mathbf{x}V^T).\label{eq:embedded_X}
\end{align*}
As a result, we can write the action of the embedding operator on computational basis elements as follows:
\begin{align*}
\mathcal{E}_V\ket{\mathbf{x}} &= \mathcal{E}_V\big(X(\mathbf{x})\ket{\mathbf{0}}\big) =   \ket{\mathbf{x}V^T}.
\end{align*}
Applying  \Cref{eq:embedded_Z,eq:embedded_X} to  \Cref{eq:embedded_stabilizers} yields the form of the check matrix of the embedded code as set out in  \Cref{eq:embedded_code}:
\begin{align}
   G_V:= \left[\begin{array}{cc|cc}
         G_X& G_X M^T & G_Z &0  \\
         0&0&M&I
    \end{array}\right].
\end{align}
\subsection{Automorphisms of Embedded Codes}\label{sec:embedded_code_auts}
We now show how to interpret the automorphisms of an embedded code as gates acting on the original codespace. The automorphisms on the embedded code space are single-qubit Cliffords and swap gates, and we show how each of these translate to the original codespace.
Let $L_V$ be a Clifford automorphism of the embedded code so that the following diagram commutes:
\begin{align*}
    \begin{CD}
    \mathcal{H}_2^{n}\otimes \mathcal{H}_2^{m} @>\mathcal{E}_V>> \mathcal{H}_2^{n+m} \\
    @V{L\otimes W}VV @VL_{V}VV \\
  \mathcal{H}_2^{n}\otimes \mathcal{H}_2^{m} @>\mathcal{E}_V>> \mathcal{H}_2^{n+m}
\end{CD}
\end{align*}
In the above diagram, the operator $W$ is some (possibly trivial) Clifford operator which preserves the group generated by $Z_i$ for $n \le i < n+m$. 
We can determine the operator $L$ on the original codespace by calculating $L\otimes W = \mathcal{E}_VL_V \mathcal{E}_V$. By construction,  $Z_j$ for $n \le j < n+m$ are stabilizers of the original codespace, and any automorphism must preserve these stabilizers. We will find $L$ by studying the action of $\mathcal{E}_VL_V \mathcal{E}_V$ on computational basis vectors and mapping the action back to a physical gate.
\subsection{Phase Operators on Embedded Codespace}\label{sec:embedded_code_S}
We consider the action of a phase operator $L_V:=S_j$ on the embedded codespace where $n \le j < n+m$. Without loss of generality, let $U_j := \textit{CNOT}_{0j} \textit{CNOT}_{1j}$ and consider the action of $U_jS_jU_j$ on a computational basis state of form $\ket{x_0,x_1,0}$:
\begin{align}
    L\otimes W\ket{x_0,x_1,0} &= U_jS_jU_j \ket{x_0,x_1,0} \\
    &= U_jS_j\ket{x_0,x_1,x_0\oplus x_1} \\
    &=  i^{x_0\oplus x_1}U_j\ket{x_0,x_1,x_0\oplus x_1}  \\
    &= i^{x_0\oplus x_1}\ket{x_0,x_1,0}.
\end{align}
The operator with this action is the phase rotation operator $L=RS_{01}=S_0S_1\textit{CZ}_{01}$ as defined in \cite{webster_transversal_2023}.
\subsection{Hadamard Operators on Embedded Codespace}\label{sec:embedded_code_H}
We now consider the action of a $H$ operator $L_V:=H_j$ on the embedded codespace where $n \le j < n+m$.  
\begin{align}
    L\otimes W\ket{x_0,x_1,0} &= U_jH_jU_j \ket{x_0,x_1,0} \\
    &= U_jH_j\ket{x_0,x_1,x_0\oplus x_1} \\
    &=  \frac{1}{\sqrt{2}}U_j\Big(\ket{x_0,x_1,0} + (-1)^{x_0\oplus x_1}\ket{x_0,x_1,1}\Big)  \\
    &= \frac{1}{\sqrt{2}}\Big(\ket{x_0,x_1,x_0\oplus x_1} + (-1)^{x_0\oplus x_1}\ket{x_0,x_1,1\oplus x_0\oplus x_1}\Big).
\end{align}
There is no operator of the form $L\otimes W$ which produces this result because it involves a superposition of states stabilized by $\pm Z_j$, so $L_V$ cannot be a valid automorphism.

\subsection{\textit{SWAP} Operators on Embedded Codespace}\label{sec:embedded_code_SWAP}
We now consider the action of a swap operator $L_V:=\text{\textit{SWAP}}_{1j}$ on the embedded codespace where $n \le j < n+m$ and both $0,1$ are in the support of $\mathbf{v}_j$.  
\begin{align}
    L\otimes W\ket{x_0,x_1,0} &= U_j\text{\textit{SWAP}}_{1j}U_j \ket{x_0,x_1,0} \\
    &= U_j\text{\textit{SWAP}}_{1j}\ket{x_0,x_1,x_0\oplus x_1} \\
    &=  U_j\ket{x_0,x_0\oplus x_1,x_1} \\
    &= \ket{x_0,x_0\oplus x_1,0}.
\end{align}
The operator on the original codespace with this action is $L = \textit{CNOT}_{01}$. 

We may also encounter a situation where the swap operator is between qubit $j$ and a qubit $k$ not in the support of $U_j$. In this case, we see that:
\begin{align}
    L\otimes W\ket{x_0,x_1,x_k,0} &= U_j\text{\textit{SWAP}}_{kj}U_j \ket{x_0,x_1,x_k,0} \\
    &= U_j\text{\textit{SWAP}}_{kj}\ket{x_0,x_1,x_k,x_0\oplus x_1} \\
    &=  U_j\ket{x_0,x_1,x_0\oplus x_1,x_k} = \ket{x_0,x_1,x_0\oplus x_1,x_0\oplus x_1\oplus x_k}.
\end{align}
For $L_V:=\text{\textit{SWAP}}_{kj}$ to be a valid automorphism and preserve the stabilizer $Z_j$, we require that $x_0\oplus x_1\oplus x_k = 0$ or $x_k = x_0\oplus x_1$, in which case $L$ acts as a logical identity on the original codespace.

\end{document}